\documentclass[sigconf]{acmart}

\AtBeginDocument{%
  \providecommand\BibTeX{{%
    \normalfont B\kern-0.5em{\scshape i\kern-0.25em b}\kern-0.8em\TeX}}}

\setcopyright{acmcopyright}
\copyrightyear{2018}
\acmYear{2018}
\acmDOI{10.1145/1122445.1122456}

\acmConference[Woodstock '18]{Woodstock '18: ACM Symposium on Neural
  Gaze Detection}{June 03--05, 2018}{Woodstock, NY}
\acmBooktitle{Woodstock '18: ACM Symposium on Neural Gaze Detection,
  June 03--05, 2018, Woodstock, NY}
\acmPrice{15.00}
\acmISBN{978-1-4503-XXXX-X/18/06}

\usepackage{_macros}
\usepackage{graphicx}
\usepackage{listings}
\usepackage{subcaption}
\graphicspath{ {./images/} }

\newcommand{\sysname}{ReSlide}

\begin{document}

\title[ReSlide]{When Constraints Limit and Inspire: Characterizing Presentation Authoring Practices for Evolving Narratives}

\author{Linxiu Zeng}
\affiliation{%
  \institution{University of Waterloo}
  \city{Waterloo}
  \state{Ontario}
  \country{Canada}
  \postcode{N2L 3G1}
}
\email{l38zeng@uwaterloo.ca}

\author{Emily Kuang}
\affiliation{%
  \institution{York University}
  \city{Toronto}
  \state{Ontario}
  \country{Canada}
  \postcode{M3J 1P3}
}
\email{ekuang@yorku.ca}

\author{Jian Zhao}
\affiliation{%
  \institution{University of Waterloo}
  \city{Waterloo}
  \state{Ontario}
  \country{Canada}}
\email{jianzhao@uwaterloo.ca}

\renewcommand{\shortauthors}{Trovato and Tobin, et al.}


\begin{abstract}
\rev{
Authoring presentation slides involves navigating contextual constraints that shape how content is structured, adapted, and reused. 
While prior work frames constraints as limitations, little is known about how presenters actively reason about them.
We conducted a formative study with ten presenters to examine how constraints emerge, are interpreted, and influence authoring decisions, leading to the Constraint-based Multi-session Presentation Authoring (CMPA) framework. 
CMPA treats \textit{time}, \textit{audience}, and \textit{communicative intent} as key constraints shaping authoring.
We instantiated CMPA in ReSlide, a research prototype for constraint-aware slide creation and reuse, and conducted two user studies on (1) single-session behaviors and (2) multi-session workflows. 
Compared to a baseline tool, ReSlide helped presenters treat constraints as active design drivers that guide narrative construction. The second study further shows how presenters flexibly reuse and adapt content across authoring cycles as constraints evolve. We then propose design implications for future constraint-aware presentation tools.
}
\end{abstract}





\maketitle

\section{Introduction}
Slides are a ubiquitous medium for presenting information effectively. 
Recognizing the need to support effective communication between presenters and audiences, many presentation textbooks decompose the entire presentation into planning, building, and delivering phases \cite{reynolds_presentation_2012, anholt_dazzle_2006, gartner-schmidt_new_2022}.
They emphasize the need for presenters to account for factors such as time limits \cite{spicer_nextslideplease_2009, lichtschlag_fly_2009, spicer_nextslideplease_2012}, audience expectations \cite{spicer_nextslideplease_2009, spicer_nextslideplease_2012}, presentation settings \cite{gartner-schmidt_new_2022}, or institutional requirements \cite{reynolds_presentation_2012}. 
We refer to these factors as \textit{contextual constraints}, which largely influence how presenters structure and deliver their presentation content, even on the same topic. 

In design and creativity research, constraints have long been viewed as double-edged: they limit users' choice but also shape and stimulate creative work. 
Although they \textit{preclude} search in some parts of the solution space, constraints can \textit{promote} search in others, pushing toward a more focused and sometimes unexpectedly novel solutions \cite{stokes_creativity_2008}.   
\rev{
Although prior works have exposed constraints in different ways, such as explicit interface elements \cite{spicer_nextslideplease_2009, lichtschlag_fly_2009, spicer_nextslideplease_2012} and static structural considerations \cite{murali_affectivespotlight_2021}, these aspects have often been treated as local capabilities or design considerations within a single presentation authoring session. 
Constraints in this sense are viewed as static requirements that the presenter has to fulfill within the local context. 
In practice, however, presenters usually start by adapting existing materials or slide decks.
}
This reality highlights slide authoring as a complex, iterative, multi-session, and constraint-driven task, and raises questions about how contextual constraints specifically are interpreted, revisited, and manipulated throughout ongoing authoring and reuse.
\rev{
Constraints have the potential to operate as ongoing drivers of authoring, shaping how presenters interpret, adapt, and reorganize content over time. 
}
However, to the best of our knowledge, this driving view of constraints has not been applied to presentation authoring. 
As a result, how constraints play a driving role in slide authoring remains underexplored. 

\begin{figure*}[tb]
  \centering
  \includegraphics[width=\linewidth]{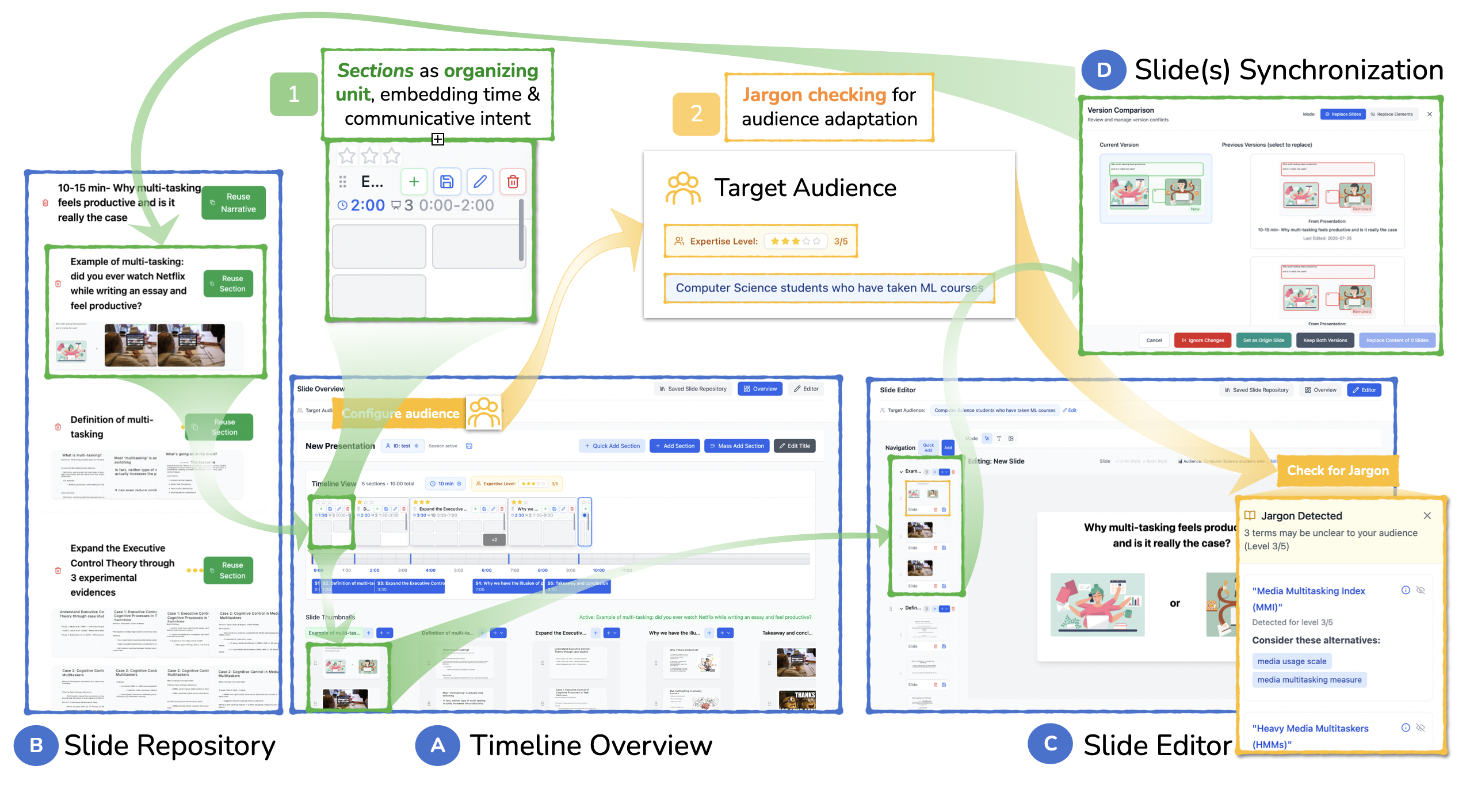}
  \vspace{-10mm}
  \caption{By instantiating our CMPA framework, ReSlide supports presenters in managing three contextual constraints—time, audience, and communicative intent—across multiple authoring cycles. (1) {\textit{Sections}} serve as intermediate units between a presentation and individual slides, grouping multiple slides around a single idea with user-defined attributes of time and emphasis. (2) {\textit{Jargon checking}} provides on-demand suggestions to adapt content to audience knowledge. Sections can be reused from the {\textit{Slide Repository}} (B), created and managed during authoring (A, C), and synced back for long-term use (D). Switching between the \textit{Timeline Overview} (A) and \textit{Slide Editor} (C) allows presenters to flexibly move across levels of detail through Sections while maintaining awareness of constraints. For long-term usage, ReSlide highlights changes in reused slides and synchronizes them with the central repository at the end of each authoring cycle.}
  \label{fig:system}
\end{figure*}

\rev{
To address this gap, we first conducted a formative study with ten participants to answer the following research question: \textbf{RQ1) How do contextual constraints guide presenters' narrative revision and editing choices during slide authoring?}
}
In this study, we mainly focused on single-presenter academic presentations, where a defined structure is required. 
Our findings surfaced recurring constraint-related patterns that confirm the guiding potential of contextual constraints and reveal how they influence slide creation practices: (1) contextual constraints act as ongoing triggers for authoring decisions; (2) edits are applied at multiple levels of granularity (e.g. section, slide, element) in slide decks; (3) reusing slide decks leaves accumulated structural traces; and (4) three main contextual constraints emerge, including time limitation, target audience's characteristics, and presenter's communicative intent. 
\rev{
We synthesized these patterns into a \textit{conceptual framework} for Constraint-based Multi-session Presentation Authoring, named \textit{CMPA}, that accounts for how evolving constraints shape presentation authoring across granularities and sessions.
}
CMPA proposes five main constructs and their relationships: ``Contextual Constraints``, ``Narrative Revision Intent``, and ``Narrative Granularity``, which capture how presenters segment and frame their stories under specific conditions, and ``Constraint-driven Authoring Behaviors`` and `Cross-session Narrative Traces``, which characterize how those conditions drive authoring actions and long-term trajectories. 
Together, these constructs articulate how presenters continuously reason about their design choices as constraints emerge, shift, and accumulate over time. 
\rev{
We thus raised two subsequent research questions: \textbf{RQ2) How does a presentation authoring tool that instantiates the CMPA framework---by making contextual constraints and narrative structure explicit---support slide authoring behavior?} and \textbf{RQ3) How do constraint-driven edits accumulate into cross-session narrative traces over time?}
}

\rev{
To investigate RQ2 and RQ3, we first instantiate CMPA through \textit{\sysname{}}, a presentation authoring tool designed to capture interconnections among the \rev{five} constructs in both single presentation and long-term authoring practice. 
}
It introduces a unit of authoring called ``Section'' (\autoref{fig:system}-1), a narrative granularity layer between the entire presentation and individual slides that gives presenters flexibility in organizing and refining ideas. 
To support reasoning about contextual constraints, \sysname{} allows presenters to explicitly specify and manipulate three representative contextual constraints identified in the formative study. 
Rather than inferring those contextual constraints automatically, we deliberately require manual specifications so that users' actions remain traceable and they retain direct control over how constraints shape presentations \cite{heer_agency_2019, roy_automation_2019}. 
Users can manually set and manage these constraints through direct manipulation during authoring (\autoref{fig:system}-C). 
Finally, a central repository (\autoref{fig:system}-B) records and demonstrates the evolution of constraint-driven adaptations across multiple authoring sessions.

\rev{
We then (1) conducted an in-lab user study with 12 experienced presenters to evaluate how \sysname{} shapes single-presentation authoring (\textbf{RQ2}), and
(2) performed an exploratory study with eight professional presenters focusing on reuse and cross-session accumulation (\textbf{RQ3}).
}
In the first study, compared to a baseline tool with basic slide authoring functionalities, \sysname{} helped presenters turn contextual constraints from peripheral checkpoints to active designing factors that guided and shaped narrative construction with greater awareness. 
\rev{
The second study further examined how presenters flexibly reuse and adapt content across multiple authoring cycles in response to evolving contextual constraints.
} 
Together, these findings consolidate the proposed framework and point to new directions for assistive presentation tools.

In summary, the main contribution of this work is threefold: 
\begin{itemize}
    \item A \textit{formative study} of constraint-driven slide authoring, deriving a \textit{conceptual framework}, CMPA, that characterizes how contextual constraints shape multi-session authoring and reuse. 
    \item A \textit{presentation authoring prototype}, \sysname{}, that instantiates the framework through sections, explicit constraint management, and a central repository for reuse. 
    \item \rev{
    A \textit{within-subjects study} and a two-session \textit{exploratory study} to empirically validate and refine the framework, as well as to surface future design implications for presentation authoring tools. 
    }
\end{itemize}

\section{Related Work}

\subsection{Conceptual Frameworks for Presentation Authoring Practice}
Prior work has proposed conceptual frameworks that rethink what a presentation is at the content and system level. 
Roels and Signer models presentation as a hypermedia graphs to support non-linear structures, content reuse, and context-aware views of presentation material \cite{roels_conceptual_2019}.
SampLe similarly encodes domain knowledge, narrative structures, media choices, and tasks to support different stages of presentation authoring \cite{falkovych_sample_2004}. 
These frameworks provides rich insights about content structures and delivery views, but they do not model how contextual constraints relate to authoring behaviors.

Other work implicitly offers practice-oriented models of presentation authoring. 
Textbooks and training materials decompose authoring into planning, building, and delivering phases, outlining procedural guidance and checklists \cite{reynolds_presentation_2012, anholt_dazzle_2006,gartner-schmidt_new_2022, schwabish_better_2017,edge_slidespace_2016,naegle_ten_2021,carter-thomas_analysing_2003}.
Systems such as SlideSpace hold similar assumptions by helping presenters plan and revise around audience and slide structure \cite{edge_slidespace_2016}. 
These work focus on procedures and tasks, but offer limited explanation of how contextual constraints operate as ongoing driving forces on narrative decisions.

Research in design and sense-making emphasizes that content creation is iterative:  
authors cycle between ``exploration'' and ``exploitation'' \cite{baker_ideas_2010, taura_signs_2011,pirolli_sensemaking_nodate}, continually reflecting on and revising ideas to evolving understanding and materials \cite{kolko_sensemaking_2010,schon_reflective_1983, klein_making_2006}. 
Narrative-focused systems operationalize this view. 
DataParticles links text, data, and visualization in a block-based editor to support iterative narratives \cite{cao_dataparticles_2023}, and work on narrative visualization distills design strategies such as ordering, messaging, transitions, and highlighting for constructing stories from data \cite{segel_narrative_2010}. 

Inspired by this view, recent tools integrate slide generation into existing workflows, generating slides from topics or documents \cite{ekart_automatically_2019, metais_smartedu_2023, maheshwari_presentations_2024, bandyopadhyay_enhancing_2024, fu_doc2ppt_2022}, or from analytics artifacts such as notebooks \cite{wang_slide4n_2023}, outlines \cite{wang_outlinespark_2024}, and interaction histories \cite{christino_knowledge-decks_2022}. 
These systems view narrative structure more than traditional one-shot generators, but still focus on converting source material into slides rather than modeling how changing contextual constraints drive ongoing narrative revisions.
Our work addresses this gap by introducing an intermediate authoring layer beyond individual slides, enabling more deliberate narrative construction with contextual awareness.

\subsection{Contextual Constraints in Design and Creative Work}
Work in creativity and design offers a richer view of constraints than simple limitations. 
Paired-constraint theories show how constraints can both restrict search in some regions and direct exploration in others \cite{alma991669233502466,stokes_creativity_2008}. 
Building on this idea, constraint-based accounts of design spaces describe design space as a conceptual space of possibilities structured by editable parameters and conditions \cite{MacLean01091991,beaudouin-lafon_prototyping_2002,heape_design_2007, dorst_creativity_2001}.
Constraints can be deliberately adjusted to guide creative work rather than simply followed as static requirements \cite{beaudouin-lafon_designing_2004}.
Similar principles have been applied in HCI research, where constraints guide interaction design \cite{fuchsberger_contextual_2014} and user-centered design \cite{norman_user_1986}, and contextual factors are treated as constraints shaping adaptive user interface \cite{dubiel_contextual_2022}.

In presentations, contextual factors such as time \cite{spicer_nextslideplease_2009, lichtschlag_fly_2009, spicer_nextslideplease_2012}, audience expectations \cite{spicer_nextslideplease_2009, spicer_nextslideplease_2012}, and presentation settings \cite{gartner-schmidt_new_2022} are recognized in different terminologies. 
Textbooks and training materials describe these as ``issues'' or ``trade-offs'' to be manually checked after slides are completed \cite{schwabish_better_2017, reynolds_presentation_2012, edge_slidespace_2016, naegle_ten_2021, carter-thomas_analysing_2003}.
Time was often operationalized as a resource to be allocated and monitored through rules of thumb and timing goals during slide preparation \cite{naegle_ten_2021, schwabish_better_2017, spicer_nextslideplease_2009, lichtschlag_fly_2009} or during delivery \cite{saket_talkzones_2014}. The \textit{knowledge gap} between presenters and audience is usually treat as a background parameter that inform tone and level of detail, rather than as elements that are revisited and reinterpreted during authoring in scientific reading and communication.
Studies of biomedical speeches showed that slides could help audiences process dense technical content with lower cognitive load \cite{dubais_use_nodate}, while corpus analyses revealed mismatches between academic vocabulary and general comprehension \cite{dang_corpus-based_2022}. 
This gap has motivated receiver-side tools to detect and simplify terminologies \cite{asthana_evaluating_2024, guo_personalized_2023,nishal_-jargonizing_2024}, provide inline explanations \cite{fok_qlarify_2024, chang_citesee_2023, lo_semantic_2023}, metaphors \cite{noauthor_towards_nodate}, or personalized visual cues for information receivers \cite{song_personalized_2025}. 
At the authoring side, prior works such as SlideSpace \cite{edge_slidespace_2016} and presentation textbooks \cite{schwabish_better_2017,reynolds_presentation_2012} encourage presenters to consider audience needs when planning a talk, but again mainly as upfront analysis checklist items.  

Across these work, contextual constraints are acknowledged as significant factors, but they tend to remain in the background as requirements. 
Instead, our framework instead treats them as driving forces for narrative design in presentation authoring, focusing on how presenters interpret and re-balance constraints across sessions during authoring.

\subsection{Multi-session Narrative Evolution and Reuse in Presentation Authoring}
Rather than starting from scratch, presenters often adapt and modify existing slides into a new presentation with a different narrative \cite{sharmin_slide-based_2012, canos_slidl_2010}. 
Manually copying and pasting slides is inefficient and error-prone \cite{sharmin_slide-based_2012}, motivating a range of tools supporting reuse across authoring sessions.

One line of work conceptualizes reuse primarily as a data-management problem.
Commercial platforms such as SlideLizard\footnote{https://slidelizard.com/en}, SlideCamp\footnote{https://slidecamp.io/}, Shufflrr\footnote{https://shufflrr.com/}, and Empower\footnote{https://www.empowersuite.com/en/} utilized a centralized library where presentation decks can be stored and shared between a central deck and local copies within organizations and collaborations \cite{kato_slide_2014}. 
Moving forward, more researches focus on content reuse in diverse scenarios \cite{cao_elastica_2024}, slide content retrieval algorithms \cite{oida_development_2018, zhang_content-based_2014, sharmin_slide-based_2012}, semantic-based slide re-adaptation \cite{spicer_nextslideplease_2009}, slide structural relationship extraction for accessibility\cite{peng_slide_2023}, and slide version tracking across multiple sessions \cite{drucker_comparing_2006,denoue_slidediff_2018}.
Beyond presentation authoring, recent works in scholarly reading and synthesis \cite{kang_synergi_2023, lo_semantic_2023} pushes reuse toward more modular and structural forms.
Overall, reuse is framed largely as a matter of managing, searching, and recombining content in repositories, while the evolving presentation narrative and contextual constraints remain implicit.

Our work instead treats multi-session reuse as part of how a presentation's narrative is re-established under changing constraints. 
Rather than focus only on assets, we emphasize how idea-level units (e.g., sections, slides, elements) are carried across presentations and re-situated as contextual constraints shift, linking reuse directly to the ways contextual constraints shape and reorganize the presentation over time. 

\section{Formative Study}
\rev{
To answer \textbf{RQ1} on how contextual constraints impact slide authoring processes, we conducted semi-structured interviews with ten experts (p1-10) to understand (1) how different contextual constraints guide narrative revision and editing choices in their workflow across multiple authoring cycles with current slide-creation tools and (2) what challenges they encountered during this process. 
}This section discusses the study in detail, followed by derived recurring themes, insights, and challenges, and ends with design goals rooted in the challenges. 

\subsection{Study Setup}
We recruited ten participants through a mailing list of a local university (5 male, 5 female, between 22 and 34 years old, average 28.3 years).
All participants had prior experience in presentation materials preparation and adaptation under different constraints, with an average of 7.2 years of presenting experience after post-secondary education.
Nine were doctoral students, and one was a postdoctoral researcher. All participants reported their discipline as human-computer interaction (HCI), related sub-fields (e.g., information visualization, AR, VR, Haptics), or interaction design. 
\rev{
Their detailed demographic information is listed in \autoref{apx:formative}. 
}

The study consisted of a one-hour interview session with open-ended questions about participants' general presentation workflows, their challenges and workarounds, and envisioned helper tools.
The study sessions were audio recorded, which were then manually transcribed, coded, and analyzed through a thematic analysis. 
This study was approved by our institutional research ethics review board.

\subsection{Recurring Themes in Current Presentation Authoring Process}
We conducted a thematic analysis of the interview transcripts. 
The transcripts iteratively coded to identify emergent patterns, and codes were then compared, merged, and refined into higher-themes, challenges, and design goals across various presentation authoring experiences under different contextual constraints.
Interpretations and theme boundaries were discussed within the research team to ensure coherence and coverage.
Our findings are organized around the recurring themes that arose throughout the process.

\subsubsection{\textbf{T1: Multi-session reuse and accumulating structural traces. }} 
\qt{Borrow[ing] slides from the [prior] presentation slides} (p8) was a common strategy across all participants. 
Several participants described starting from an existing deck, then selectively keep those could be reused, readapted, or repurposed in response for the new context (p7, p8, p9), while initiating corresponding authoring behaviors such as adding new content (p1, p3, p6, p7, p8), removing content no longer useful (p10), expanding existing contents (p1, p8), or shortening them (p6) to accomodate changes in audience groups (p8) or the storyline (p7). 
Participants sometimes explicitly mentioned their preference in reusing the \qt{newest version}, because \qt{some minor changes may not be reflected in the oldest version} (p7). 
Even without a fully formed storyline, this reuse process helped presenters craft and tune a new narrative to the current context, with reuse decisions becoming narrative decisions guided by contextual constraints. Some participants referred to a long-term slide repository as \qt{the main deck} (p8), or a form of \qt{version controlling} (p10), describing it as the most comprehensive version, containing multiple slide versions tailored for different contextual constraints. 
Upon completion of a slide, some participants deliberately updated the main branch (p8), synchronizing changes from the latest presentation back into this comprehensive repository. 
Across these practices, the slide content was not static; rather, slides continuously evolved over multiple sessions and carried traces of older versions in their structure.

However, this way of managing content and reuse was often inefficient and left multiple duplicates, resulting in an increasingly complex structure over time.
Manually synchronizing the newer slide version back to the main deck was difficult: \textit{when I make changes to the specific versions, I hope it applies to the main deck as well} (p8). 
As a result, some participants defaulted to using the latest version even when sometimes \qt{the storyline may not align that well} (p7), because intermediate changes in \qt{middle} presentations were only minor and not fully merged back with other versions. 
Over multiple sessions, these constraint-driven edits accumulated into duplicated sections, fragmented versions, and partially edited copies that were difficult track and manage.

\subsubsection{\textbf{T2: Contextual constraints and triggers for authoring decisions.}} 
Participants consistently reported that they started an authoring session by considering contextual constraints and how they might shape the storyline. 
Common contextual constraints were \textit{audience}, \textit{communicative intent} (or the \textit{goal} of the presentation), and \textit{time}. 
Yet once they started working on specific ideas or slides, they often lost track of these factors.
Instead, constraints moved in and out of attention and re-entered at particular moments, each time triggering concrete authoring decisions. 

At the beginning of a new slide-authoring session, audience and communicative intent were the primary factors for shaping this emerging storyline.
All participants emphasized that \qt{having your audience in mind is very important} (p4), recognizing doing a presentation as inherently an audience-oriented task \cite{edge_slidespace_2016}. 
Presenters needed to tailor the presentation to a specific audience's needs (p1, p6), especially their knowledge background (p2). 
Failing to consider these could lead to comprehension challenges (p2) and misalignment between the intended message and the audience's understanding (p3).  
Additionally, communicative intent influenced storyline crafting (p7), using it to decide which parts to prioritize or downplay.
However, the cognitive load of balancing content creation with adaptation often leads to either overlooking the intended adaptations altogether or postponing them to later iterations---such as simplifying language, reducing jargon or technical words (p2, p3), and proactively tailoring content to the audience (p2, p5, p8). 
Some presenters emphasized the importance of aligning the slides with an overarching communicative intent. 

Time followed a slightly different pattern.  
It was usually known upfront as a hard constraint, but it did not strongly shape story-building in the earliest stages. 
Instead, it re-entered attention once a draft storyline, or slide deck, was in place, when presenters were ready to move one detail level up and refine pacing and weight. 
Participants mentioned relying on heuristics such as the \qt{two slide per minute} rule (p9) and repeatedly rehearsed in later stages to trim content (p1, p4, p7, p9, p10). 
Yet even with those strategies, they still found it hard to balance the amount of content with the overall story flow (p9). 
This was caused by a lack of overview of the storyline, which resulted in redundant work throughout the authoring cycle (p7).

\subsubsection{\textbf{T3: Operating across narrative granularities.}}
Participants did not treat individual slides as the only unit of authoring.
Instead, they repeatedly moved between different narrative granularities when structuring and revising their content, from the overall presentations to sections, to slides, and to elements within a slide. 
Several participants expressed the need to gain \qt{a good overview of the storyline} (p7) and to manage both high-level and lower-level transitions (p2, p7).
Relying on a purely slide-based approach made this difficult. 
Achieving this required carefully \qt{go[ing] through all the slides and examining content} (p1), which caused \qt{a lot of redundant work} (p7) and drifting away from the main point (p10).
Some mentioned the one-idea-per-slide strategy (p2, p6), but it quickly became unmanageable as slide counts grew large (p6), which further emphasized the need to reason beyond individual slides.

Many authoring actions were carried out at the section level or in groups of slides. 
Maintaining coherence required attention to distributing and transitioning across sections (p4).  
As one participant noted, \qt{I need to make sure that the transition between sections is consistent} (p9). 
Additionally, others found the content distribution should correctly signal importance, as \qt{[inappropriate allocation] will cause the audience to lose their sense} (p3). 
This required participants to inspect the pacing and corresponding emphasis at the section level, not just within or across single slides.

At the same time, participants need flexible control. 
Some operations are \qt{slide-based} (p9), like reordering or duplicating slides, while others are \qt{element-based} (p7, p10), such as reusing or revising a specific figure, text, or layout. 
Content reuse acts as a strategy to save time with on-demand action across multiple granularities, enabling flexible reuse across groups of slides (p7, p8, p9), a single slide that \qt{conveys one piece of [context independent] information} (p7, p9), or a single element (p10), depending on the current context. 
Together, these practices highlight that presenters routinely operate across multiple narrative granularities, but current tools primarily focus on slide-level view management and leave the transition to manual effort.

\section{Conceptual Framework: Constraint-driven Multi-Session Presentation Authoring}

\subsection{Problem Space and Design Gap}
Our framework focuses on a single-presenter, slide-based academic presentation in which slide decks are reused across multiple authoring sessions. 
We denote one ``authoring session'' as the period during which presenters work on one presentation slide deck.
In common presentation authoring practice, rather than creating a deck once and discarding it, presenters repeatedly revisit, reuse, adapt, and repurpose it across different contexts. 

Prior work and existing tools addressed important aspects of this space, but in isolation. 
Time-management systems assisted presenters to better assign duration during slide preparation \cite{spicer_nextslideplease_2009,lichtschlag_fly_2009} or during delivery \cite{saket_talkzones_2014},  
presentation textbooks and prior works emphasized audience awareness \cite{schwabish_better_2017, reynolds_presentation_2012, murali_affectivespotlight_2021},  
commercial tools offered centralized libraries for presenters to store and share slide copies as well as track version histories within an organization, and 
retrieval algorithms were proposed to help users more efficiently locate relevant slides\cite{oida_development_2018, zhang_content-based_2014,sharmin_slide-based_2012}. 
Conceptual frameworks for presentation, such as Roels and Signer's content framework \cite{roels_conceptual_2019}, defined rich information-layer structures for reuse, collaboration, and context-aware delivery. 

\begin{figure*}[tb]
    \centering
    \includegraphics[width=0.9\linewidth]{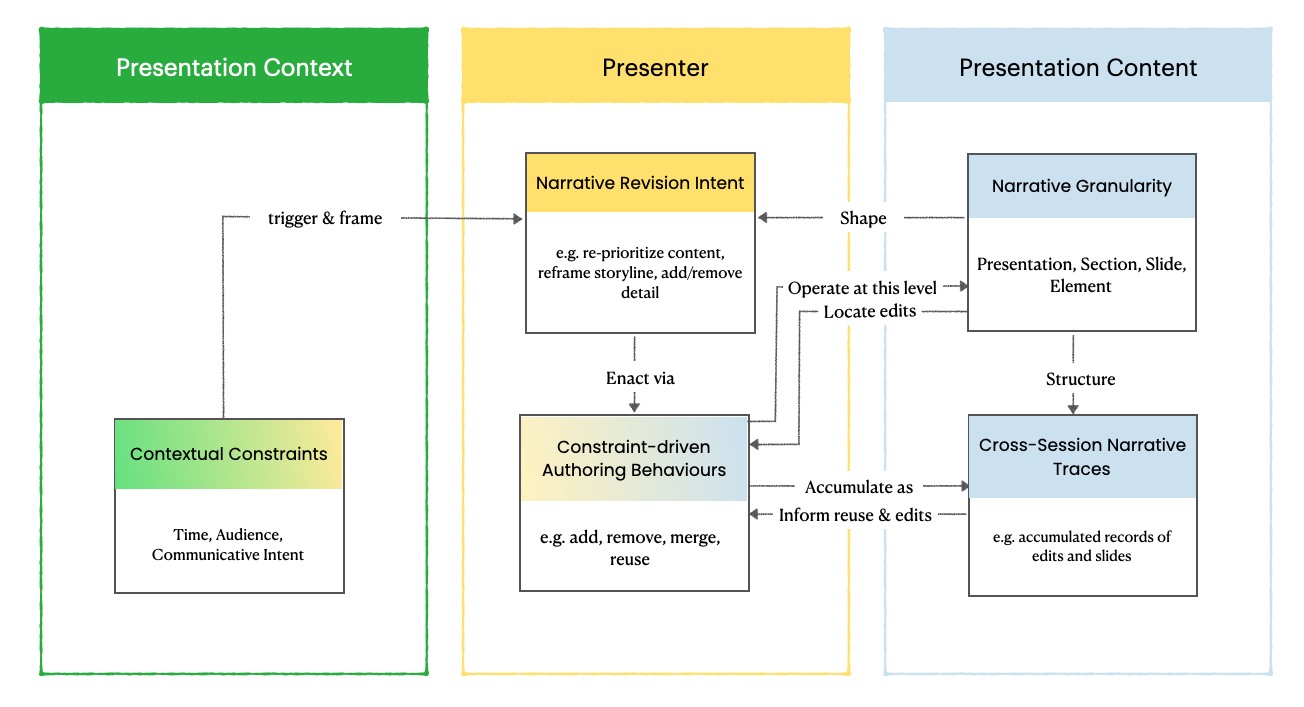}
    \vspace{-4mm}
    \caption{The Constraint-based Multi-session Presentation Authoring (CMPA) framework is modelled with the Presentation Context, Presenter, and the Presentation Content as the three main entities. The Presentation Context refers to the broader presentation setting that provides potential constraints (e.g., event, venue, schedule). The diagram highlights five core constructs with associated examples. Arrow edges point towards the influenced factor (see \autoref{subsec:core-constructs}).  }
    \label{fig:framework}
    \centering
\end{figure*}

However, none of these approaches model how contextual constraints are used as ongoing driving forces during multiple authoring sessions,  
how these forces are expressed through different authoring behaviors across different narrative granularities (sections, slides, and elements),  
or how those operations accumulate as structural traces that shape subsequent sessions. 
This leaves a missing explanatory layer: how presenters repeatedly re-interpret constraints as triggers for deciding what to keep, change, create, or reuse. 
Unlike prior work that primarily models content structures and context-aware delivery, our framework models how constraints function as driving forces that shape reuse and authoring decisions.  

\begin{figure*}[tb]
    \centering
    \includegraphics[width=\linewidth]{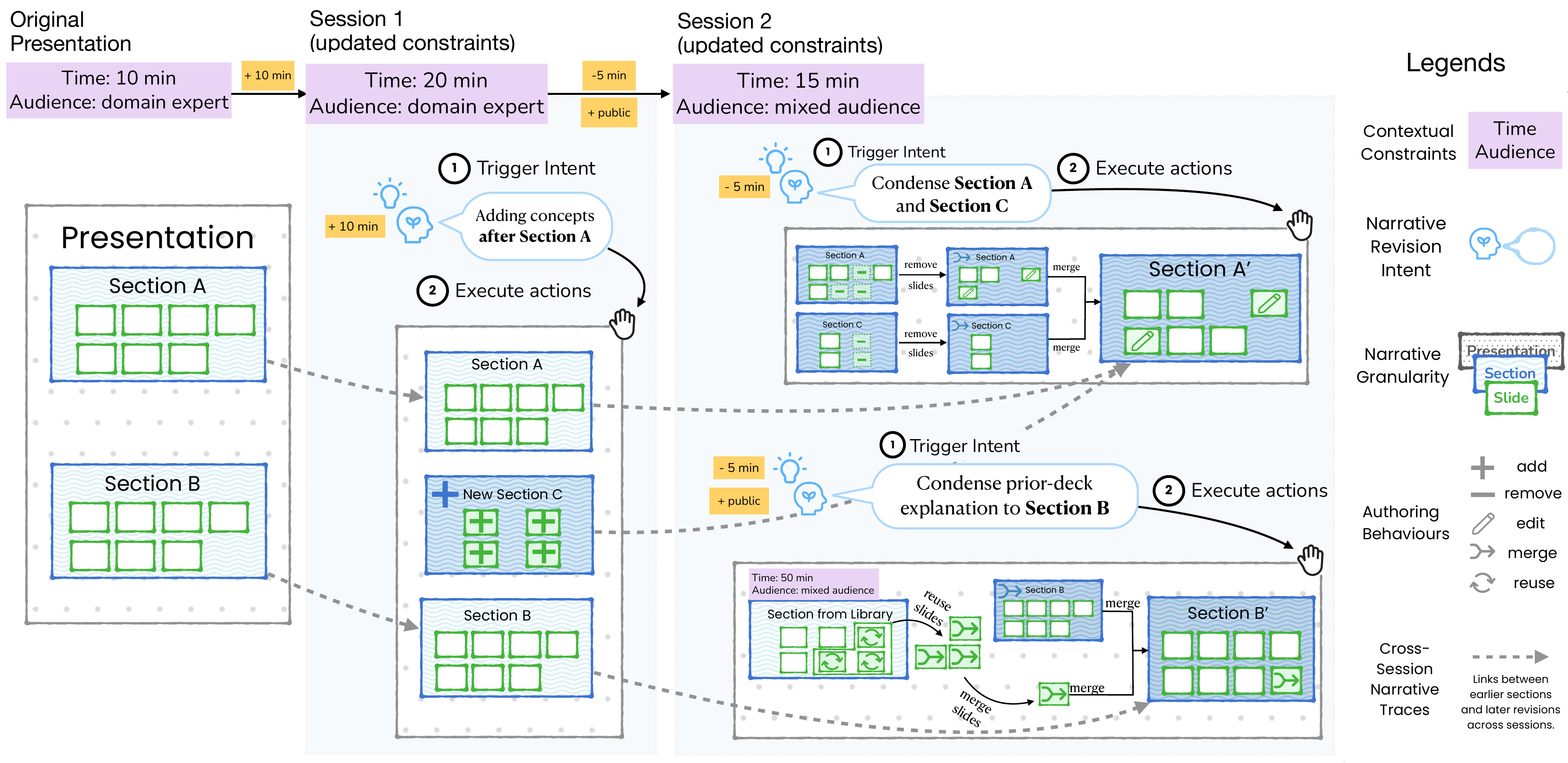}
    \vspace{-2mm}
    \caption{This figure provides an example of how CMPA explains multi-session slide authoring and reuse. 
    As contextual constraints (time, audience) change between Session~1 and Session~2, presenters (1) update their narrative revision intent (blue thought bubbles), and (2) act at different narrative granularities using concrete authoring behaviors.  
    The right panel summarizes the framework's core constructs with explanations or illustrative instances. 
    \rev{
      In the example, Session 1 adds a new adjacent Section C when more time becomes available, and Session 2 then condenses and merges content from Section A and C into a revised Section A' as earlier changes are carried forward and reshaped under updated constraints.
    }  
    Section B is carried over in Session 1 and later refined in Session 2, where slides are merged and adapted, including material reused from a prior deck, illustrating how constraint-driven edits and reuse accumulate across sessions.}
    \label{fig:framework-example}
    \centering
\end{figure*}

\subsection{Core Constructs and Mechanisms}
\label{subsec:core-constructs}
Based on the recurring themes in formative study, we derived the \textit{Constraint-based Multi-session Presentation Authoring (CMPA)} framework (\autoref{fig:framework}), targeting constraint-based multi-session presentation authoring that treats contextual constraint as a core design factor. 
CMPA proposes five main constructs and their relationships. 
In short, the change of Contextual Constraints triggers Narrative Revision Intent, where presenters mentally decides the current narrative and content should change.
Narrative Granularity captures the unit of focus at which this intent is enacted.
Concrete actions are carried out through Constraint-driven Authoring Behaviors, and the resulting content changes leave traces that accumulate into trajectories across multiple sessions.
We then define each construct with empirical groundings and theoretical support, and then integrate its relationships into one explanatory model.

\subsubsection{Contextual Constraints}
We define this construct as the situational conditions that presenters consider when deciding what to keep, change, create, or reuse during authoring.
These conditions are not fixed inherent properties of the existing materials; rather, they are interpreted by the presenters in a given situation. 
They may be specified upfront or may emerge as the presenter refines the presentation; they may shift and be reinterpreted within and across multiple authoring sessions. 
In CMPA, we focus on three representative constraints identified in the formative study: time, audience, and communicative intent. 
Participants repeatedly referred to the change of those constraints as reasons for editing decisions (e.g., \qt{cut to fit time,} \qt{reframe for a mixed audience}). 
Constraints can be applied at different granularity levels, shaping which unit becomes the primary focus of revision.
In \autoref{fig:framework-example}, constraints are shown on a per-session basis, and their changes are highlighted in yellow as triggers for subsequent authoring behaviors.

\subsubsection{Narrative Revision Intent}
This construct describes the presenter's current decision about how the story should change in response to the contextual changes before committing to specific slide edits, such as what to emphasize, reorganize, reduce, or introduce.
It provides an explanatory bridge between contextual constraints and concrete authoring actions. 
In CMPA, rather than directly picking specific edits, contextual constraints first drive presenters to update their narrative revision intent, such as re-framing communicative intent, re-prioritizing content, adjusting storyline order, or re-wording certain expressions. 
This intent is then externalized through authoring behaviors at a chosen granularity of content intervention.
Participants often began the session by reviewing contextual constraints and how they might shape the storyline before selecting existing presentation content that could be reused, readapted, or repurposed (p7, p8, p9).
For example, if the audience group changed from domain experts to a general audience, participants were prone to add more background information at the beginning and adjust existing explanations to the same content, rather than changing the storyline. 
Those actions reflect the same narrative intention of adding weights of those parts in their storyline. 
We demonstrate user intention as speech bubbles in \autoref{fig:framework-example}.

\subsubsection{Narrative Granularity}
This construct describes the organizing unit that presenters frame and apply revisions. 
We observed that narrative granularity spans four levels in our data from high to low: presentation, section, slide, and element. 
Participants alternated between restructuring at the section level and refining at the slide level, depending on what they were trying to satisfy. 
Thus, granularity is a property of the slide deck and also reflects the presenters' current authoring focus and the coherent unit for decision-making. 
While constraints can apply across multiple levels, Narrative Granularity captures which level is currently in focus. 
In \autoref{fig:framework-example}, we demonstrate granularity using different-sized boxes to indicate the location of edits (e.g., which units are added/removed/merged) and structural decomposition.

\subsubsection{Constraint-driven Authoring Behaviors}
Constraint-driven authoring behaviors are observable editing actions that presenters use to enact narrative revision intents triggered by contextual constraints.
In this work, we focus on concrete operations that change narrative structure or content allocation, such as add, remove, merge, and reuse. 
As observed in the formative study, participants used these actions to align with the changing constraints (e.g., selectively delete slides no longer useful (p10), or add new content for broader audience groups(p1, p3, p6, p7, p8)).
These behaviors can be applied to multiple narrative granularities. 
For example, merging may involve combining slides within a section or merging multiple sections, both reflecting the same narrative intention of reducing the weight of certain parts in the storyline when available presentation time is reduced.
By externalizing decisions, these behaviors leave structural traces that can be tracked across sessions, including changes in section composition or slide allocation.
\autoref{fig:framework-example} illustrates that a single constraint shift (\ie presentation time reduces 5 minutes) may trigger merging in both section-level and slide-level. 

\subsubsection{Cross-Session Narrative Traces}
This construct captures how edits and reuse accumulate over multiple authoring sessions, producing trajectories of narrative restructuring rather than isolated changes. 
Participants described organizing multiple versions of the same content into a \qt{main repository} for future use (p8, p9), indicating that earlier revisions often constrain or enable later revisions in new contexts. 
Cross-session narrative traces emphasize the accumulation and path dependence of how content evolves in relation to shifting contextual constraints over time. 
It summarizes these structural traces produced by authoring behaviors and explains how these traces influence subsequent revision and reuse.
In \autoref{fig:framework-example}, cross-session narrative traces are illustrated by the reused section from a prior deck being adapted into Section B$'$, and by the chained mini-versions of Section A that accumulate into the final Section A$'$.

\subsubsection{Constructs Relationships and Mechanisms}
Together, the constructs describe a recurring process of constraint-driven multi-session authoring.
Shift in Contextual Constraints prompts two aspects of Narrative Revision Intent: whether to revise and at what granularity to intervene (presentation, section, slide, or element) within and across sessions. 
This intent captures the mental reasoning of how the story should change and which level should be targeted. 
Narrative granularity then specifies the unit of change, and contextual constraints provide the criteria that guide what should be kept, modified, created, or deleted at that unit. 
Then, presenters enact those decisions through constraint-driven behaviors (add, remove, merge, etc.), potentially applying multiple behaviors in response to the same constraint shift. 
Over time, the resulting revisions accumulate into structural traces that form a temporal trajectory of the slide deck, where trajectory refers to the sequence of sections or slide versions created across sessions as constraints change.
This trajectory reflects the reasoning process of how presenters use contextual constraints as a thinking guide to continuously refine presentations. 
The resulting traces become the starting point for subsequent authoring sessions, shaping how the narrative continues to evolve.

\subsection{Scope and Boundaries}
CMPA is illustrated through three contextual constraints identified in the formative study: time, audience, and communicative intent. 
These representative constraints cover only a subset of all possible contextual constraints.
We do not claim these three constraints exhaust the space; rather, they provide representative anchors for illustrating how constraints function as triggers in multi-session reuse. 
This framework focuses on authoring rather than live-delivery, multi-user collaboration, presentations involving multiple agents, done in industrial settings, and not done within a large company. 
It does not account for highly improvised presentations nor fully automated slide generation.

\section{\sysname{}}

\subsection{Design Goals}
Based on the conceptual framework, we derived the following key design goals inform the design of our presentation authoring tool, \sysname{}, which instantiates the framework with concrete features.

\textbf{DG1: Enable visibility of contextual constraints along presentation authoring.} 
Authoring tools should encourage and assist presenters to externalize and monitor relevant constraints by making them explicit at appropriate narrative units, surfacing conflicts that triggers new revisions, tying constraints to narrative units, and providing subtle reminders to support continued alignment.

\textbf{DG2: Facilitate structured narrative construction based on idea pieces and relationships. }
Tools should support fluid navigation across narrative granularities, allowing presenters to structure ideas at multiple narrative granularities, group slides, and assign emphasis level. 
Such structured mappings help presenters maintain coherence between evolving slides and the intended narrative flow.

\textbf{DG3: Support flexible reuse and synchronizatiaon of contents at different granularities. }
Tools should be able to assist content reuse and synchronize across narrative granularities. 
They should preserve provenance, support decisions when updating reused content, and allow changes to accumulate into trajectories across sessions.  
This integrate both a single authoring session and long-term content management into a natural slide authoring workflow.

Taking these guidelines in mind, we introduce \sysname{}, a presentation authoring tool that aligns contextual constraints with evolving presentation narratives. 

\begin{figure}[tb]
    \centering
    \includegraphics[width=\linewidth]{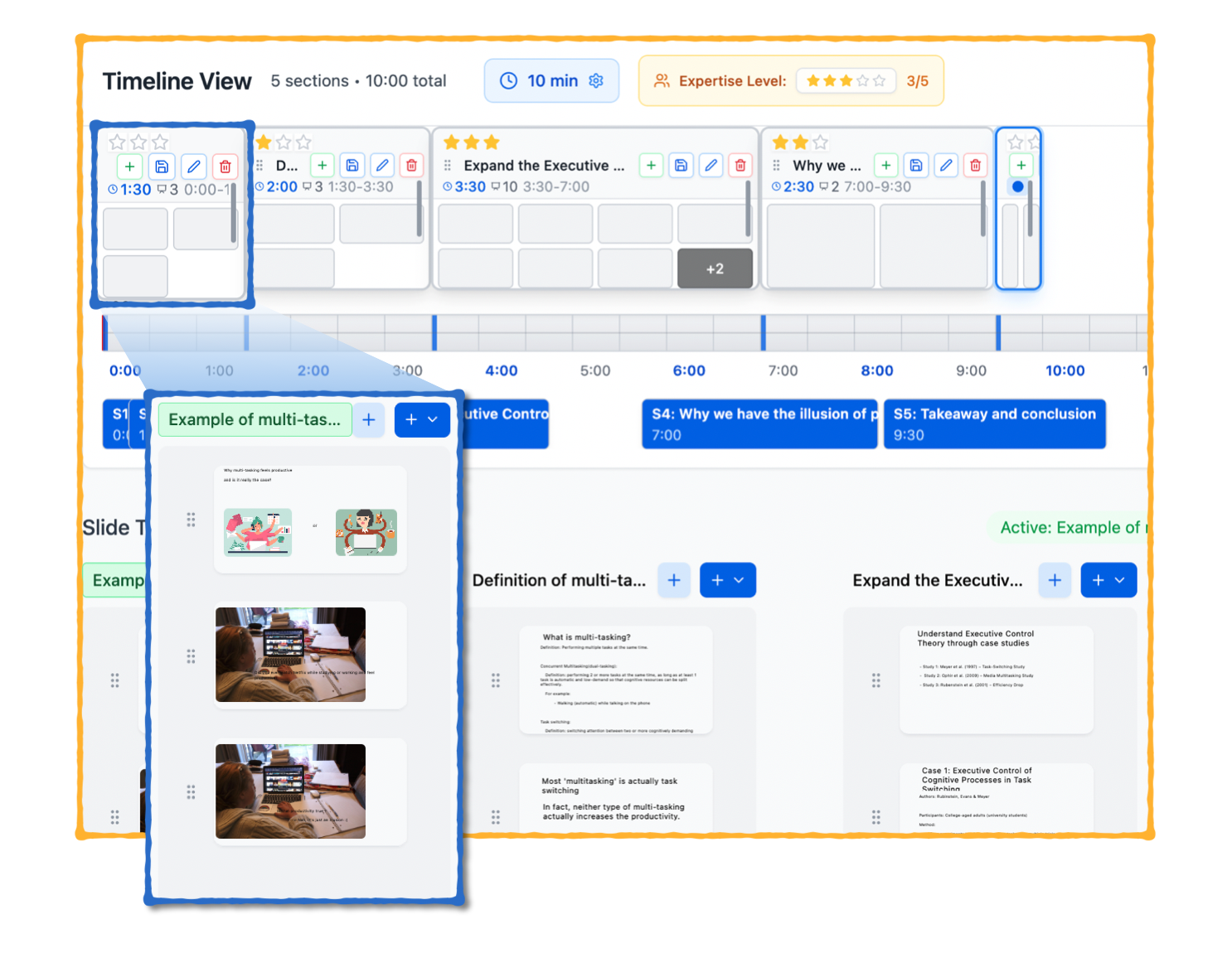}
    \vspace{-7mm}
    \caption{Each Section is visualized as a vertical bar below the timeline, displaying its associated slide thumbnails.}
    \label{fig:section-vertical}
    \centering
\end{figure}

\subsection{System Overview}
\sysname{} was implemented with a React frontend and Node.js backend.
The frontend user interface consists of three views: 1) a \textit{Timeline Overview} (\autoref{fig:system}A), a \textit{Slide Repository} (\autoref{fig:system}B), and a \textit{Slide Editor} (\autoref{fig:system}C). 
\sysname{} introduces Section (\autoref{fig:system}-[1]) as the core unit of authoring, which is a group of consecutive slides representing a semantically coherent idea. 
This structure gives presenters more flexibility in managing content while focusing on building ideas (\textbf{DG2}). 
The Timeline Overview presents the narrative structure of the presentation, showing all created sections with accumulated time (\textbf{DG1}, \textbf{DG2}).
Below the timeline, each section is visualized as a vertical bar (\autoref{fig:section-vertical}), with its associated slides grouped inside (\textbf{DG2}). 
The Slide Repository lists reusable resources at multiple granularities (presentation, section, slide), enabling content retrieval and reuse (\textbf{DG3}).
This promotes long-term flexible usage by offering four options to synchronize reused slides with prior versions (\autoref{fig:system}D).
The Slide Editor provides a workspace organized by sections and incorporates jargon checking (\autoref{fig:system}-[2]) to adapt content to target audiences, supporting both constraint awareness and narrative construction (\textbf{DG1}, \textbf{DG2}). 
Together, these features enhance presenters' awareness of contextual constraints, support explicit narrative revision intent at the section and slide levels, and integrate these decisions into a natural narrative construction workflow across multiple presentation authoring cycles.

The backend of \sysname{} comprises two main components: an \textit{API layer} and a \textit{Jargon Detection Module}. The API layer manages the storage and retrieval of reusable presentation content by communicating with MongoDB for slide content and AWS S3 for image data. 
The Jargon Detection Module analyzes the audience information to identify potential jargon in the current slide content. 
This component embeds GPT-4o-mini via the OpenAI API, applying prompt engineering to generate simplified alternatives tailored to specified audience. 
The processed outputs are relayed to the frontend components for visualization and interaction. 

\subsection{Define Unit of Authoring with Section}
\begin{figure}[tb]
    \centering
    \includegraphics[width=\linewidth]{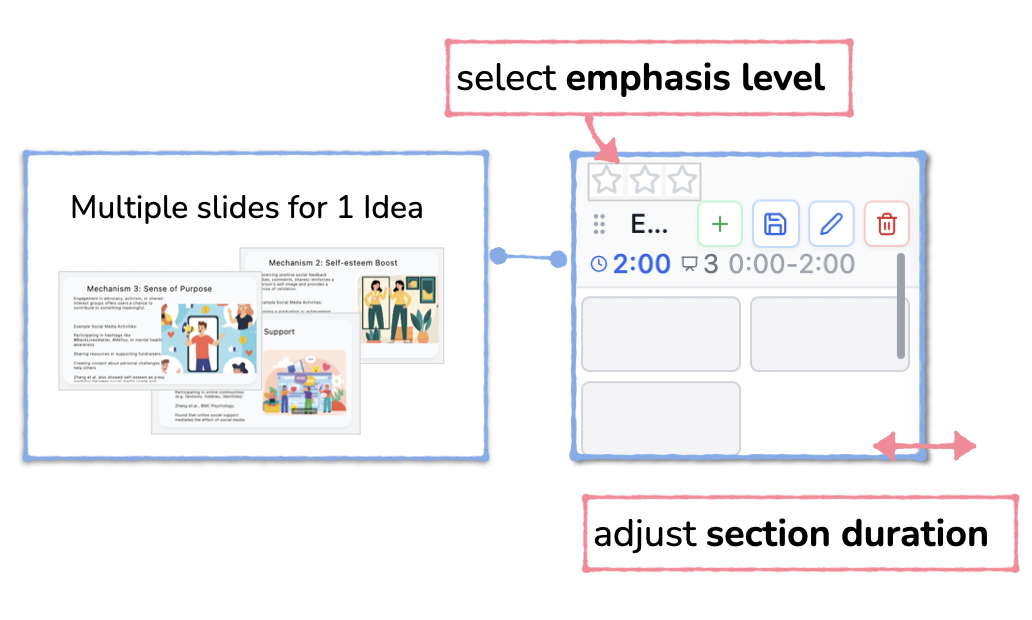}
    \vspace{-8mm}
    \caption{Sections serve as a core unit of authoring in \sysname{}. Each Section represents one idea and can contain multiple slides, with time duration and emphasis level set as parameters. }
    \label{fig:section}
    \centering
\end{figure}

To help presenters effectively manage contextual constraints (\textbf{DG1}) and organize content while constructing a storyline (\textbf{DG2}), we introduce the idea of Section. 
This is inspired by agenda or overview slide, a common practice where presenters provide the audience with a roadmap of the presentation. 
\sysname{}'s Section aligns with the presenters' natural mental model of how a presentation flows, where one section represents a single idea and may contain multiple slides (\autoref{fig:section}). 
\sysname{} proposes a section-first workflow, in which the users first create sections and then adds slides within them.
This mirrors how presenters mentally group content into narrative units and directly supports the Narrative Granularity in our framework. 

\begin{figure}[tb]
    \centering
    \includegraphics[width=1.1\linewidth]{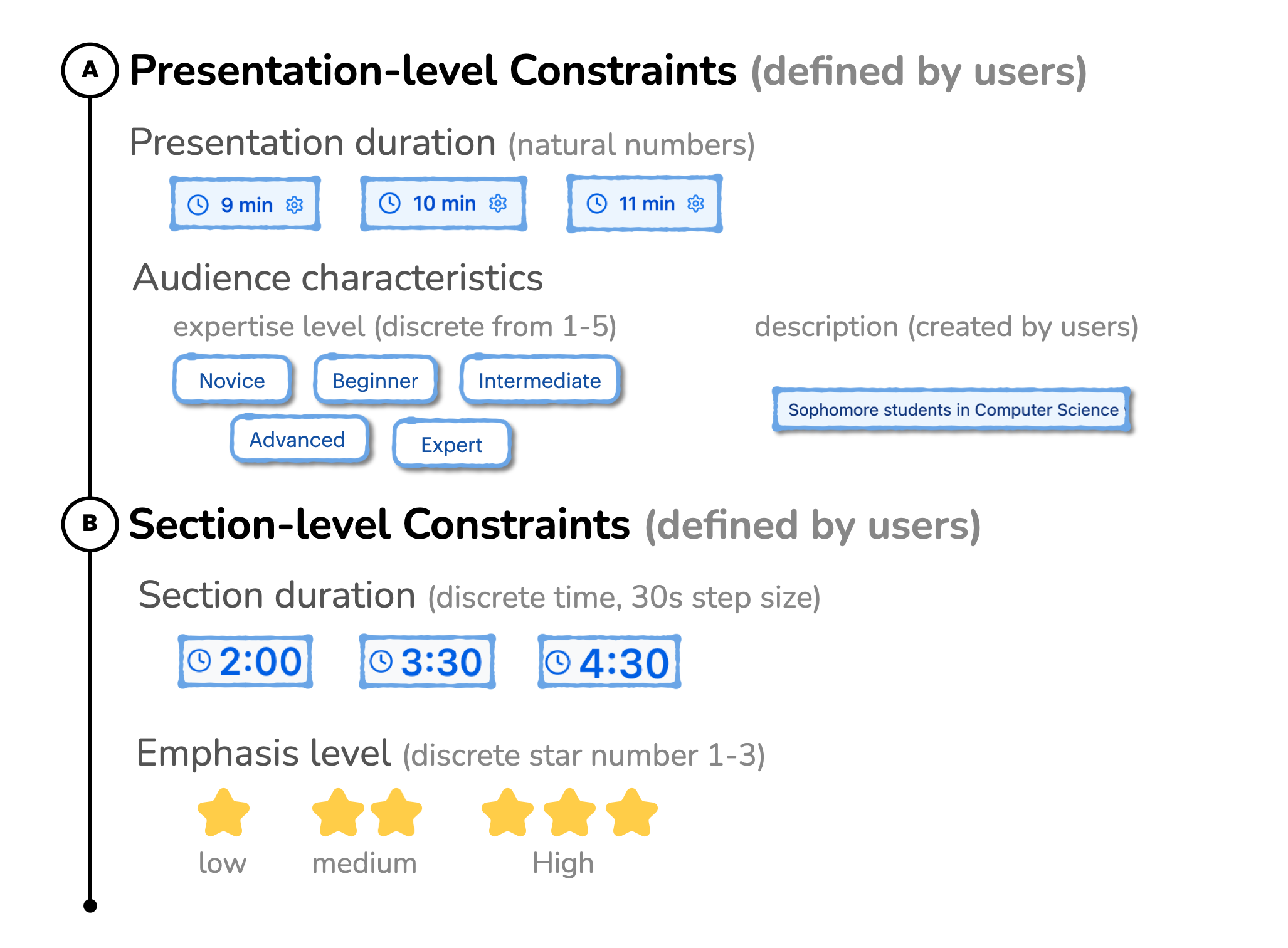}
    \vspace{-6mm}
    \caption{User-defined constraints in \sysname{}. At the presentation level (A), presenters specify overall duration and audience characteristics. At the section level (B), presenters define section duration and emphasis level. }
    \label{fig:constraints}
    \centering
\end{figure}

Presentations usually come with a certain time limit, but our formative study found that considering only the overall limit is too vague, often causing later stage iterations. 
When the user creates a new presentation, \sysname{} requires them to set the total time (\autoref{fig:constraints}).
Additionally, when creating a section, the user can set its time duration, estimating how long this part may occupy in the presentation; otherwise a default of 2-minute is applied. 

Presenters may not know exact section durations early on, but usually have a sense of how important each idea is to the presentation or audience, which relates to the audience and the communicative intent constraints (\autoref{fig:constraints}). 
To better assist the users in mapping their intended storyline to the actual content, \sysname{} allows them to set an emphasis level for each section.
There are three levels of emphasis---low, medium, high---indicating how important that section is over the entire narrative. 

\begin{figure}[tb]
    \centering
    \includegraphics[width=\linewidth]{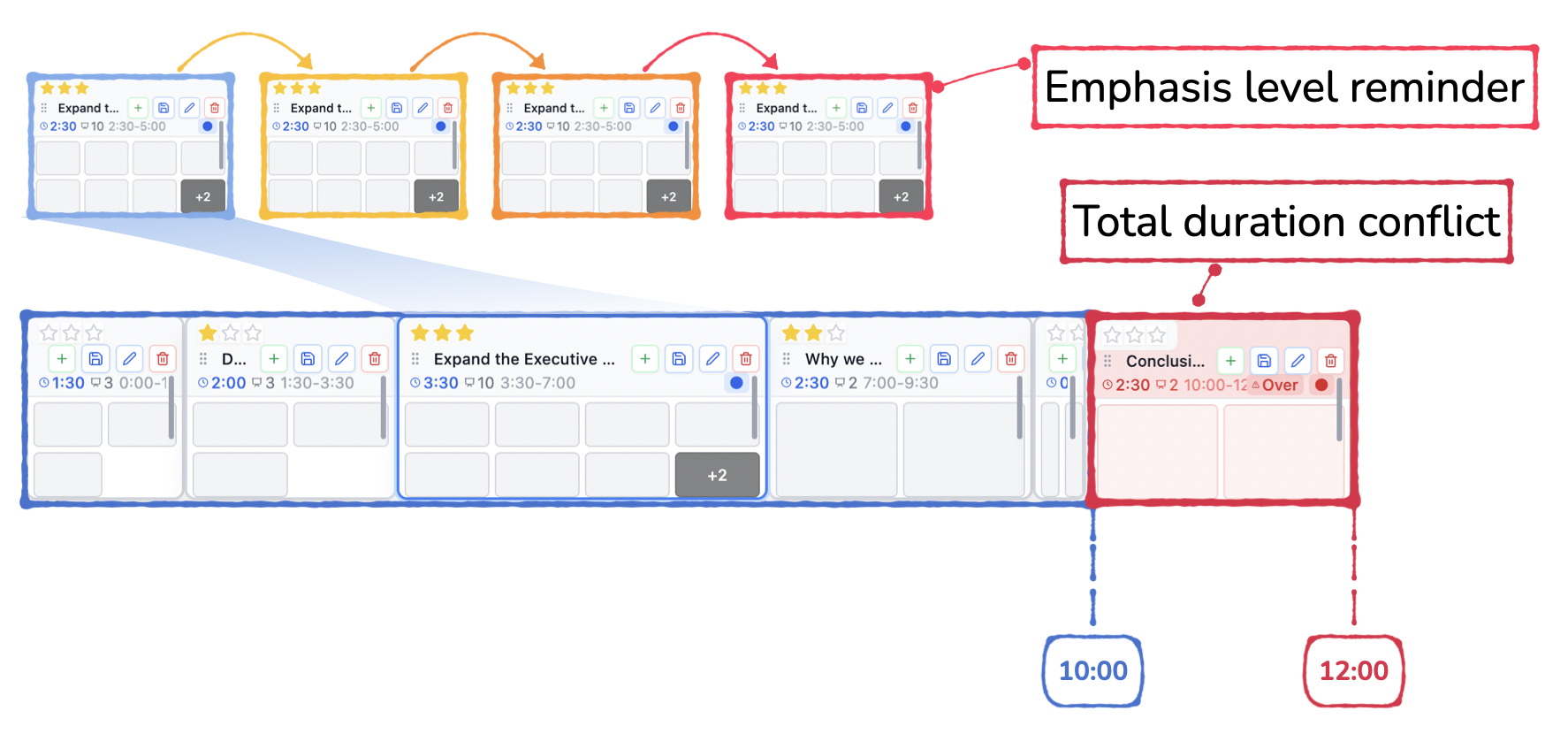}
    \vspace{-5mm}
    \caption{The Timeline Overview highlights constraint conflicts. Sections with mismatched emphasis levels and time durations are color-coded: blue indicates no conflicts, while conflicts ranging from yellow to red. If the total duration exceeds the time limit, the exceeding section(s) are marked in dark red.}
    \label{fig:conflictReminder}
    \centering
\end{figure}

Section time and emphasis level work together. 
Conflicts areise when a high-priority section receives less than than a low-priority section (e.g., a high-priority Key Result section is assigned 2 minutes while a low-priority Conclusion is allocated 4 minutes).
Such sections are highlighted as soft indicators, reminding presenters to reconsider the balance between timing and narrative flow (\autoref{fig:conflictReminder}).   
There are three levels of indication represented by three colors, ranging from yellow to red, indicating an increasing conflict; blue means no conflict. 
A yellow mark signals low conflict when an important section receives about the same time as a less important section; an orange mark indicates medium conflict when an important section receives 75\% of the less important ones; a red mark shows high conflict when an important section receives only 50\% or less than the less important ones.
These indicators ensure that time allocation and communicative intent remain aligned by encouraging presenters to revisit their narrative revision intent when mismatches arise.

\begin{figure*}[tb]
    \centering
    \includegraphics[width=\linewidth]{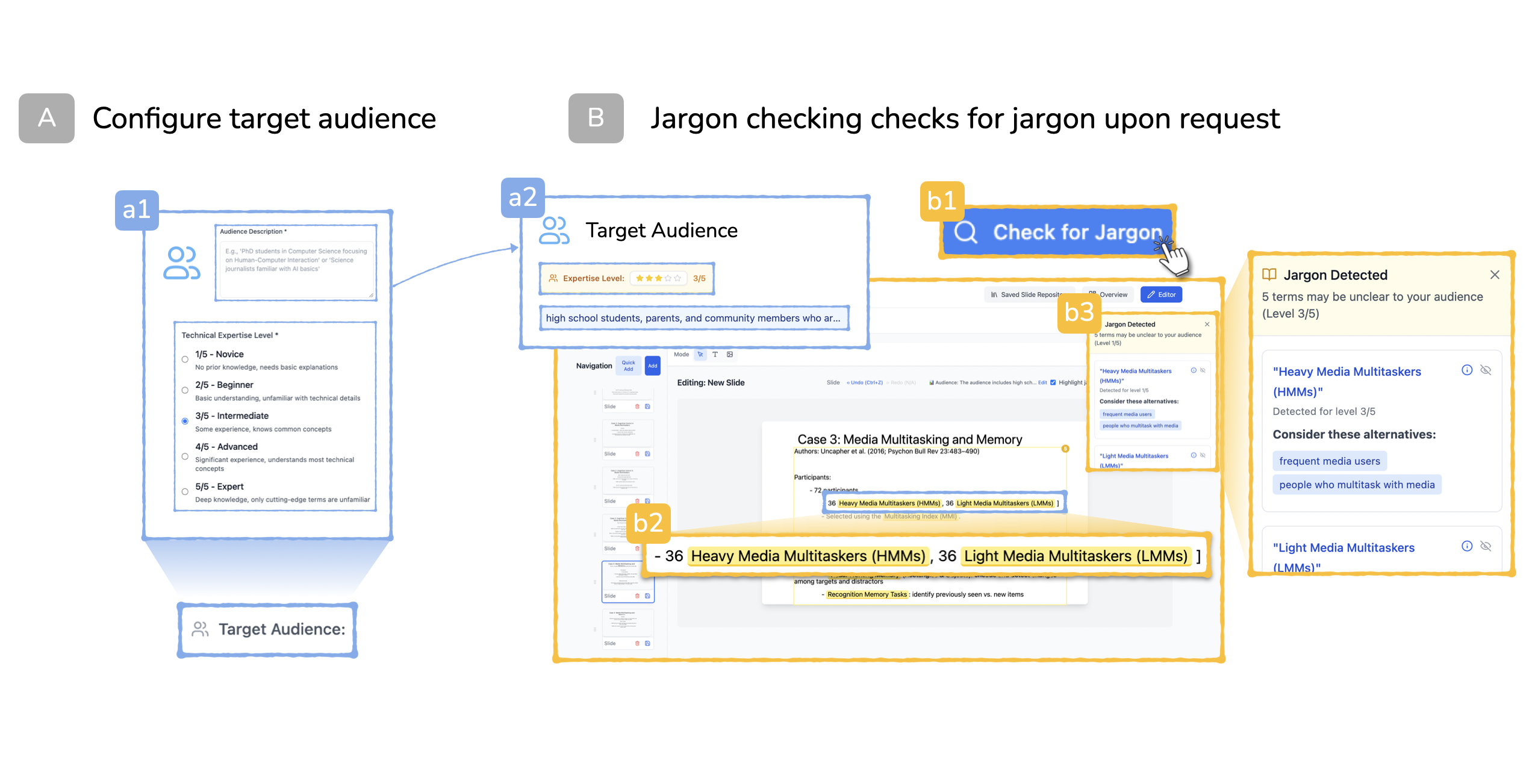}
    \vspace{-16mm}
    \caption{At the start of an authoring cycle, the presenter configures the \textit{target audience} (A) by selecting expertise level and providing a brief description, which is then displayed in the Editor View (a1-a2). When the jargon checker is triggered (b1), the system scans the current slide for terms that may exceed the audience's modeled expertise. If jargon is detected, the system 1) highlights the term in its original location (b2) and 2) lists all flagged terms in a Side Panel (b3), showing each term's expression with an on-demand definition and two suggested simpler alternatives tailored to the modeled audience.}
    \label{fig:jargon-checking}
    \centering
\end{figure*}

\subsection{Support Narrative Construction with Timeline and Jargon Detection}

Based on authoring units exhibited as sections, \sysname{} supports a visible constraint management (\textbf{DG1}) and a structured narrative building (\textbf{DG2}) with the interactive Timeline Overview and jargon checking feature.

To accommodate different workflows, \sysname{} allows sections to be added individually, in bulk, or as placeholders for rapid outlining. 
Once created, sections are displayed in the Timeline Overview (\autoref{fig:system}A), providing a high-level context of the narrative structure while surfacing potential conflicts.  
Presenters can refine the narrative by dragging to reorder or adjusting their duration directly. 
Slides can also be organized and moved across different sections by drag-and-drop. 
Together, the timeline visualization allows presenters to manage detailed slide content without losing control of the overall structure. 
Following the presenters' mental model, the Timeline Overview (\autoref{fig:system}A) and the Slide Editor (\autoref{fig:system}C) effectively externalize the presenters' mental workload while supporting a clearer storyline development experience. 
Switching between the Timeline Overview and the Slide Editor allows presenters to alternate between high-level narrative revision and detailed content editing, without losing sights of how these decisions relate to the contextual constraints.

We model audience-related contextual constraints explicitly and use them to drive a jargon detection feature that supports narrative revision intent at the wording and example level. 
Upon creating a new presentation, \sysname{} requires users to describe the intended audience (\autoref{fig:constraints}). 
\sysname{} models the audience in two ways (\autoref{fig:jargon-checking}A-a1): 
\begin{itemize}
    \item \textit{Expertise level}: Presenters can categorize their audience on a five-point scale, ranging from 1 (Novice) to 5 (Expert). A higher number indicates greater prior knowledge about the topic.  
    \item \textit{Audience description}: Presenters can write a short description to capture extra information not represented by the expertise level. 
\end{itemize}
These settings guide the jargon detection in \sysname{}.
The lower the audience expertise the presenter sets, the more detailed and strict the system checks for potentially difficult expressions, languages, etc., correspondingly. 
\rev{
Upon receiving the audience profile,
}
\sysname{} first expands the user's input audience description and expertise rating into a detailed profile that includes background, inferred expertise, known concepts, likely jargon areas, and domain context. 
This profile is then used in a second analysis of slide content, detecting jargon and suggesting tailored alternatives or definitions while considering user characteristics and presentation context. 
Both steps prompts are implemented using ChatGPT API calls, and the prompts are listed in \autoref{apx:jargonPrompts}.

The jargon checker can be applied at the slide level and triggered on demand (\autoref{fig:jargon-checking}B). 
When activated, the system checks the current slide and reports each potential jargon by: 1) highlighting the jargon in its original position (\autoref{fig:jargon-checking}B-b2), and 2) displaying the expression, the on-demand definition, and two suggested simpler alternatives tailored to the modeled audience on a side panel (\autoref{fig:jargon-checking}B-b3).
\rev{
\sysname{} presents these as suggestions rather than automatic modifications, and presenters must manually revise the slide content.
}
Presenters are also allowed to hide individual words or all flagged terms if they find the suggestions not useful, or if they have already resolved the suggestions.

\subsection{Enable Multi-Layered Presentation Reuse and Synchronization with Central Repository}

\begin{figure}[tb]
    \centering
    \includegraphics[width=\linewidth]{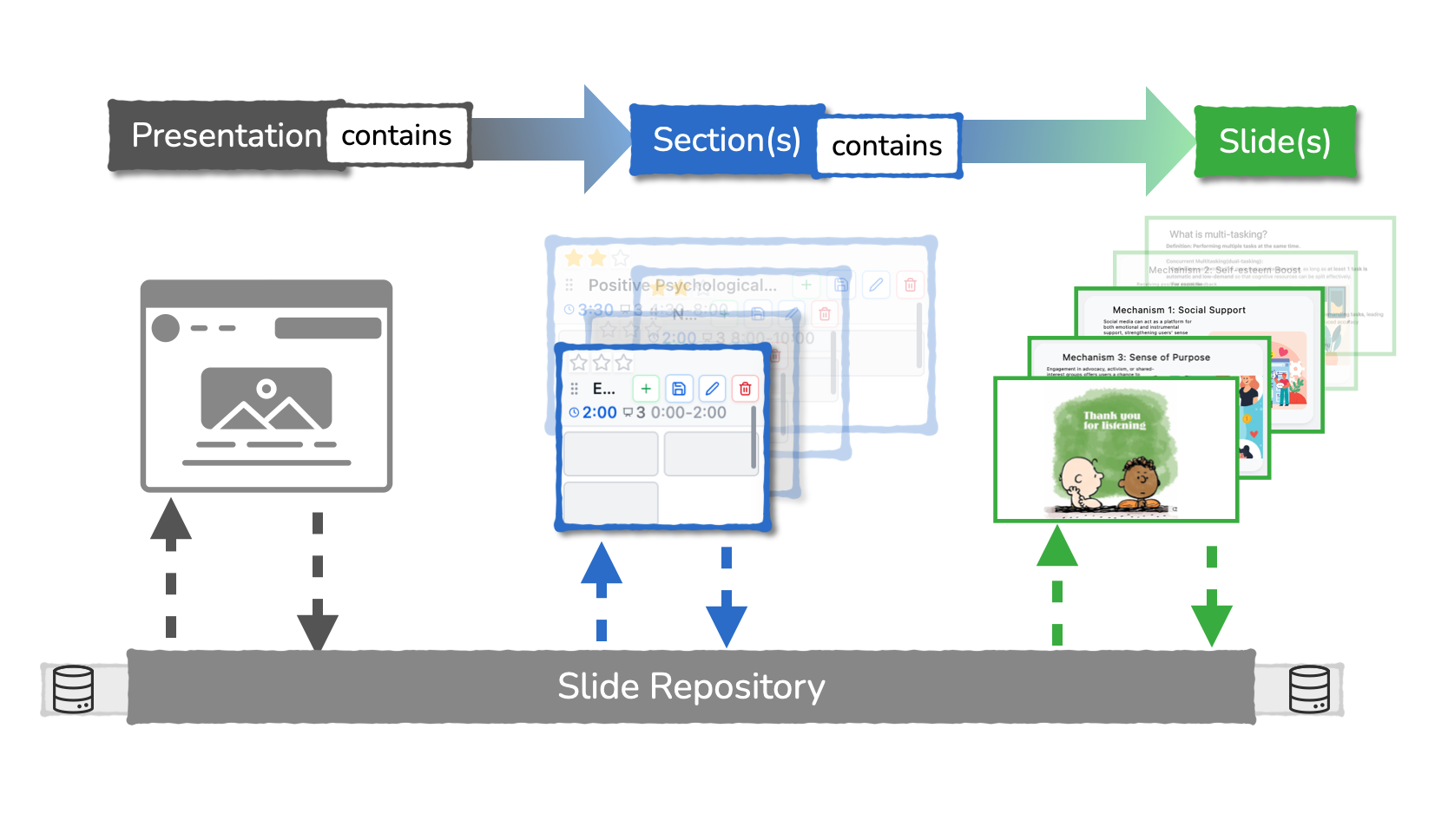}
    \vspace{-9mm}
    \caption{\sysname{} organizes content in a hierarchical structure where a presentation consists of sections, and each section consists of multiple slides. This structure allows content in multiple granularities to be stored in and retrieved from the central Slide Repository. }
    \label{fig:reuseUnit}
    \centering
\end{figure}

To support long-term efficiency and reuse (\textbf{DG3}), \sysname{} introduces a central Slide Repository. 
Content can be managed at multiple granularities---presentation, section, and slide---where each has consistent saving and reusing behaviors  (\autoref{fig:reuseUnit}). 
This reflects how presenters naturally manage their presentation structure while providing flexibility across different knowledge scales. 
Saving the higher-level structure preserves its connection with the lower-level structure.
By storing content at multiple granularities and maintaining lineage across versions, the Slide Repository materializes cross-session narrative traces: as constraints change over time, the trajectories of section and slides evolve as well.

At the presentation level, the complete deck, including all sections with their corresponding metadata (\eg, duration, emphasis), is stored in the repository.
A saved presentation can be re-imported into the workspace as its full narrative, enabling presenters to adapt to new contextual constraints.
Similarly, at the section level, sections and their metadata can be stored, and restored as a copy to a new workspace. Thus, presenters can then adjust the newly-imported section to fit a new storyline.
Finally, at the slide level, each slide is linked to all of its previous versions. Presenters can update individual components (e.g., text boxes, figures) on a slide.
Upon saving, the system checks for changes and asks the presenter to synchronize slide versions via a Slide Comparison Panel (\autoref{fig:system}D). 
There are four options to resolve the synchronization across versions: 
\begin{itemize}
\item \textit{Ignore Changes}: discard all changes and keep the prior version.
\item \textit{Set as Origin Slide}: save the new slide as the starting point of a new lineage.
\item \textit{Keep Both Versions}: save the new slide while keeping lineage. 
\item \textit{Replace Content}: overwrite the selected prior version(s) with the new slide or with updated elements.
\end{itemize}
Later, an individual slide can be retrieved through a keyword search. Selected slides should be placed into a chosen section. Hovering over slides in the repository shows an enlarged preview that assists with browsing and selection. 
These options allow presenters to express how the current revision should relate to prior versions under the specific contextual constraints.

\begin{figure*}[tb]
    \centering
    \includegraphics[width=0.9\textwidth]{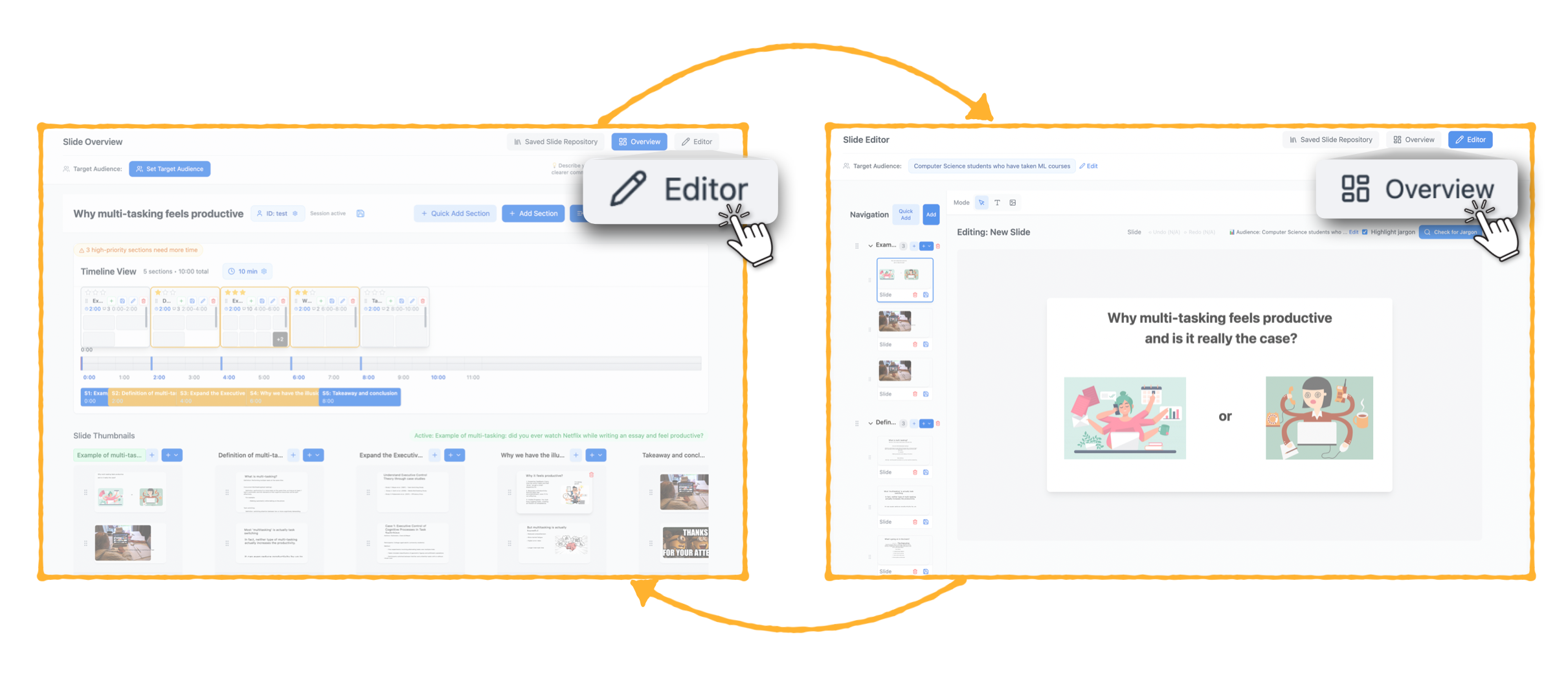}
    \vspace{-7mm}
    \caption{The workflow involves switching between the Timeline Overview and the Slide Editor, enabling users to alternate between high-level structuring and detailed content editing.}
    \label{fig:two-view}
    \centering
\end{figure*}

\subsection{Example Workflow}
Below we present an example workflow to demonstrate some of the features described above. 
Alice, a psychology PhD student, is preparing a ten-minute talk on the topic ``Why multitasking feels productive,'' a topic relevant to one of her publications. 
Her audience includes middle to high school students, parents, and community members.
Alice begins by launching \sysname{}, setting her total presentation time (10 minute) (\autoref{fig:constraints}A), entering the audience description---high school students, parents, and community members who are curious about this topic but may have a range of backgrounds in it (\autoref{fig:jargon-checking} a1-a2)---with expertise level to three, reflecting their limited prior knowledge. 

Then, she outlines the storyline by creating sections, assigning emphasis levels, and adjusting their duration (\autoref{fig:section}).
For example, she includes an Introduction and Conclusion with no emphasis, as these are not central to her main point.
Between them, she adds three body sections (\autoref{fig:section-vertical}): one defining multitasking (low emphasis, 2 minutes), one explaining the illusion of increased productivity (high emphasis, 3.5 minutes), and one discussing daily implications (medium emphasis, 2.5 minutes). 

To construct her slides efficiently, Alice strategically reuses the Introduction section from a prior presentation to peers that is stored in her repository (\autoref{fig:system}B). 
She also incorporates several content slides from that presentation into her explanation section. 
While drafting, she activates the jargon checking, which flags the term ``Heavy Media Multitaskers (HMMs)'' and suggests two alternatives: ``frequent media users'' and ``people who multitask with media'' (\autoref{fig:jargon-checking} b1-b3). 
She thus integrates these revisions to ensure content clarity for her audience. 
Once she is done with a section, she switches back to the Timeline Overview to ensure she stayed aligned with the constraints (\autoref{fig:two-view}). 
Once all slides are completed, Alice saves and synchronizes (\autoref{fig:system}D) them back to the central slide repository (\autoref{fig:system}B), making them available for future reuse.
This example illustrates how \sysname{} supports constraint-driven authoring: Alice first configures contextual constraints, then revises her narrative at the section level, enacts these decisions through concrete authoring behaviors, and finally records the resulting narrative traces in the repository for future reuse.

\section{User Studies}
We conducted two in-lab user studies using \sysname{} as an instantiation of the CMPA framework to examine how the framework manifests in practice and to gain insights into it.  
\rev{
Study 1 employed a within-subjects design, comparing \sysname{} with a baseline system to probe how contextual constraints shape presentation authoring, addressing \textbf{RQ2} on how a tool that instantiates the CMPA framework shapes slide authoring.
Study 2 was an exploratory study where participants used \sysname{} across two authoring scenarios, which simulated multi-session authoring practice, allowing us to investigate \textbf{RQ3} on how constraint-driven edits accumulate on cross-session narrative traces.
}

\subsection{General Study Setup}
We conducted both studies in person after receiving approval from the university's ethics review board. 
Study sessions were audio and video recorded, lasted around 100-130 minutes, during which participants completed assigned tasks, post-task questionnaires, and a semi-structured interview. 
Participants were compensated with \$40 CAD.
For our presentation authoring tasks, we created a pool of four topics that were accessible to most people, preventing potential presumptions about participants' backgrounds: 
\begin{itemize}
    \item Understand everyday inflation behind rising grocery prices?
    \item How does social media impact mental health?
    \item How AI and personalized suggestions shape what you see online?
    \item Why multi-tasking feels productive, and is it really the case?
\end{itemize}
\rev{
The topics were chosen to balance technical and social domains and were iteratively refined by the research team to be similar in scope, feasible for participants with diverse backgrounds, and not overly abstract or likely to distract presenters from developing a structured presentation. 
}

\subsection{Study 1: Comparative Study with Single-Session Authoring}

We compared \sysname{} with a baseline system that mimics current slide authoring tools (e.g., Google Slides, Microsoft PowerPoint, Apple Keynote). 
To ensure fairness between the two systems and minimize bias from participants’ varying familiarity with existing tools, we built a customized baseline that provided only basic slide authoring functionalities (i.e., text and image) and slide reuse.
The authoring functions were identical in both systems.
In the Baseline, participants could reuse slides via copying and pasting, which is similar to how slide reuse works in commercial presentation authoring tools.

\subsubsection{Participants}
We recruited 12 experienced presenters (three males, nine females; between 23 and 33 years old, average 26 years) via social media and mailing lists. 
\rev{
All participants held at least a bachelor's degree; 11 were enrolled in postgraduate programs, including four pursuing PhDs.
}
Participants were asked to self-report their slide-preparation and presentation skills through a set of screening questions. 
All participants were familiar with at least 1 slide creation tool. 
Ten out of 12 participants have received training in public speaking or presentation skills, while the 2 who have not received training reported frequent presentation experiences in the past 3 months or had adequate experience of presenting to mixed audiences. 
The screening questions and the detailed demographic information are listed in \autoref{apx:evaluation}.

\subsubsection{Task} 
Participants were asked to create a presentation based on a given scenario under a specified topic selected from the topic pool.
During the task, participants had access to: a short scenario description, some supplementary materials to help them familiarize themselves with the topic, a slide deck previously created for a different scenario on the same topic, and space for notes or drafting. 
We encouraged presenters to adapt the provided slide deck while creating the new storyline.
However, if the participants felt an extra burden while adapting that slide deck into the current slide, they could also not reuse any of the reusable slides.

\subsubsection{Procedure}
After signing the consent form, participants were given an overview of the study procedure, duration, and data collection details. 
The study was conducted in person, with participants accessing both systems via a research computer. 
\rev{
All study data were de-identified and handled securely during storage, transfer, and sharing.
No personal information was used in the study tasks. 
}
A brief tutorial, followed by a brief training session that included a set of warm-up tasks, going through all the features, was provided to demonstrate the use of both systems. 
Participants could familiarize themselves with the tools before starting the tasks.

Participants were instructed to pick two topics from the topic pool that they were more familiar with.
Then they used the two systems (\sysname{} and Baseline) to perform two presentation authoring tasks, each corresponding to a different topic they selected. 
The order of the systems was counter-balanced across participants. 
Participants were given 30 minutes to complete one task.
A countdown timer was provided, and the researcher also gave time indications every 10 minutes after the task started. 
After each task, participants rated the system usability, cognitive load, intent expression, and slide authoring experience with the system on a 7-point Likert Scale. 
Upon completing two tasks, a semi-structured interview was conducted to gather feedback on participants' experiences.
Participants were encouraged to share general comments on any aspects of the study and then prompted on specific aspects, including interface usability, pros and cons of the systems, as well as prior difficulties and strategies tackling the contextual constraints and reuse slides. 
They were also asked about the differences between systems' impact on their approaches to solving their raised strategies and suggestions for improvement. 
Observations noted by the researcher during the session were also discussed.

\subsection{Study 2: Exploratory Study with Multi-Session Authoring}
This study was designed to explore the multi-session authoring practice of \sysname{}. 
For this purpose, we created two scenarios: (1) a public talk at a local library (time: 5-10 min, audience: general audience, communicative intent: explaining concept and application), and (2) an academic conference presentation (time: 10-15 min, audience: domain experts, communicative intent: discussing technical details).

\subsubsection{Participants}
We recruited eight experienced presenters (four males, four females; between 25 and 39 years old, average 30 years) via social media and mailing lists. 
All participants were familiar with at least one presentation authoring tool. 
\rev{
Seven were PhD students, and one held a bachelor's degree.
}
They were more experienced presenters who satisfied at least two of the following: Have done 6–10 presentations in the past 3 months; Have had formal training before; Have been presenting for 5+ years since starting post-secondary; Have usually presented to a large audience (e.g., more than 50 people). 
Their detailed demographic information is listed in \autoref{apx:evaluation}. 

\subsubsection{Task} 
Participants created slides on one topic of their choice from the topic pool.
Unlike Study 1, participants were instructed to create presentations for the above two scenarios on the same topic, enabling the reuse of their own slides in the second round.
During the first task, participants had access to: a short scenario description, some supplementary materials to help them be familiar with the topic, and space for notes or drafting. 
In the second task, besides the above items, participants also had access to the slides they created.

\subsubsection{Procedure}
The beginning of the procedure for Study 2 was the same as that for Study 1, until the actual study tasks.
Participants were instructed to build two presentation slides on the same topic, each under different constraints as described above.
They were encouraged to think-aloud during the presentation authoring.
After each authoring task, participants rated their experience with the system on a 7-point Likert Scale. 
Upon completion of two tasks, we conducted a semi-structured interview to gather feedback on participants' experiences with \sysname{}.
The questions in the interview were similar to those in Study 1, and observation notes were discussed.

\subsection{Data Collection and Analysis}
Evaluating presentation quality is inherently subjective, especially when participants are instructed to focus on storyline construction. 
Therefore, we combined standardized surveys, custom self-report ratings, interaction logs, and interviews to examine how our framework's constructs are manifested in practice. 
To assess usability and cognitive effort while creating a presentation, we leveraged the UMUX-Lite and NASA-TLX in Study 1 and analyzed the results with the Wilcoxon signed-rank test. 
\rev{
To capture participants' internal reasoning about constraints, narrative structures, and reuse, we designed a set of custom questions. 
}
Those questions fell into three categories (constraints, coherence, and reuse), capturing participants' subjective impression while doing tasks (see \autoref{apx:evaluation}). 
In both studies, we logged participants' interaction data to understand how they managed their constraints and structured their slides. 
We analyzed the interview data using thematic analysis, following similar procedures to the formative study with both inductive and deductive approaches. 
We open-coded the transcribed interviewees' responses, employing affinity diagramming to sort the initial codes. 
Through iterative discussions and the organization of these codes, we identified a number of recurring patterns and themes. 

\section{Results}

\begin{figure*}[tb]
    \centering
    \includegraphics[width=0.9\linewidth]{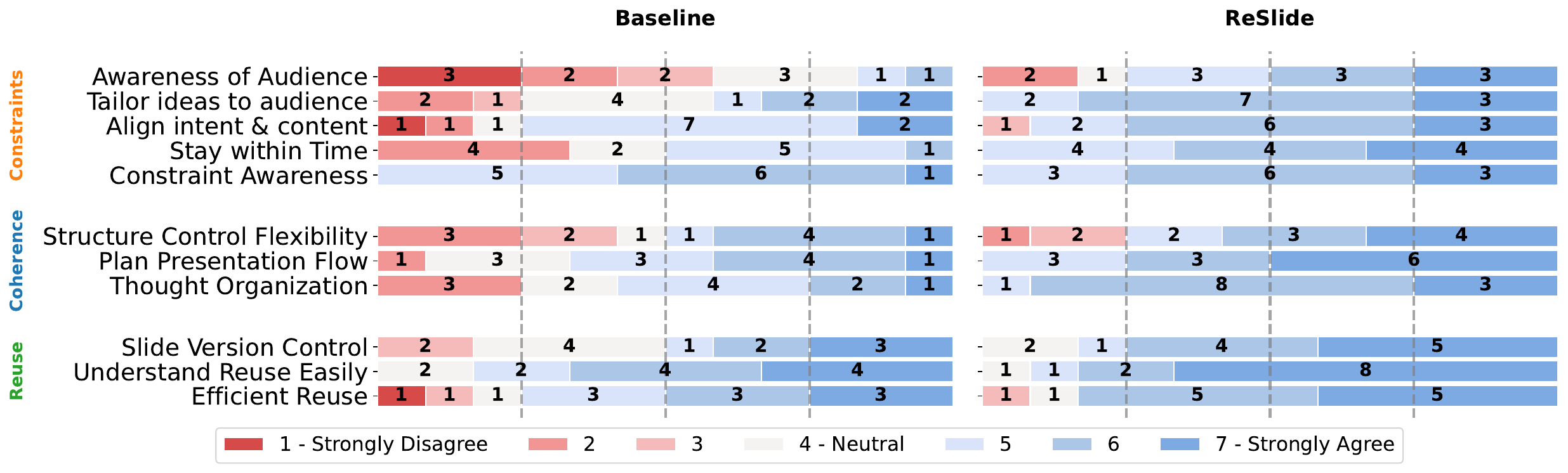}
    \vspace{-1mm}
    \caption{Participants' responses in Study 1 for comparing presentation authoring experiences with Baseline and \sysname{}.}
    \label{fig:likert_group}
\end{figure*}

\rev{
In this section, we report the results from the two user studies. 
We begin with the questionnaire results in Study 1 (\autoref{sec:quant}), which provide an overview of perceived workload, usability, and support for constraints, coherence, and reuse. 
We then aggregate and present the qualitative results from both studies (\autoref{sec:theme1}--\autoref{sec:theme3}), as both investigate presenters' authoring behaviours through the CMPA-driven tool, \sysname{}. Discussing the qualitative results together allows us to better highlight recurring themes and patterns across contexts. In Sepcific, Study 1 examines how contextual constraints and narrative granularities shape presenters' reasoning path, storyline intervention, and edit choices (\textbf{RQ2}), and Study 2 focuses on how versioning decisions accumulate into cross-session narrative traces (\textbf{RQ3}). 
}
Participants in Study 1 are referred to as \textit{c1-12}, and those in Study 2 are referred to as \textit{e1-8}.




\subsection{Questionnaire Results from Study 1} \label{sec:quant}
Although seven of 12 participants reported lower perceived workload on the NASA-TLX for \sysname{}, the difference between Baseline and \sysname{} ($Mdn_B = 3.33$ vs. $Mdn_R = 3.42$, $p= 0.414$) was not significant. This suggests that our system did not add cognitive or physical burden while making presentations, even though it offered many new features. 
The system usability scores computed from UMUX-LITE were significantly greater ($p = 0.021^*$) for \sysname{} ($Mdn_R = 6$) compared to Baseline ($Mdn_B = 5.25$). 
Participants consistently reported significantly better results with \sysname{} on all subjective metrics (\autoref{fig:likert_group}), from Contextual Constraints ($Mdn_B = 4.4$ vs. $Mdn_R = 5.9$, $p = 0.002$, $r = 0.815$), Narrative Coherence ($Mdn_B = 4.16$ vs. $Mdn_R = 6.16$, $p = 0.014$, $r = 0.706$), to Reuse ($Mdn_B = 5.17$ vs. $Mdn_R = 6.67$, $p= 0.020$, $r = 0.667$).

\subsection{Contextual Constraints as Driving Forces on Narrative Reasoning and Edits (RQ2)} \label{sec:theme1}

Baseline led to a freeform authoring workflow similar to participants' prior experience with existing tools, which prioritized slide content over contextual constraints.
Time, audience, and communicative intent were mostly kept in their heads. 
They relied on their own judgment to stay aligned, which often led to inefficiencies or to ignoring the constraints altogether during delivery. 
With \sysname{}, participants instead treated constraints as something to set and revisit. 
They described \qt{setting [them] as input for the interface} (e7), assigning initial time estimates to each section in advance (c12), and later performing a final check to ensure alignment (e1). 
Although some found it difficult to estimate time or importance early on, entering constraints upfront effectively gave them a concrete starting point for shaping the storyline (c7, e1, e7).

\begin{figure*}[t]
    \centering
    \includegraphics[width=\linewidth]{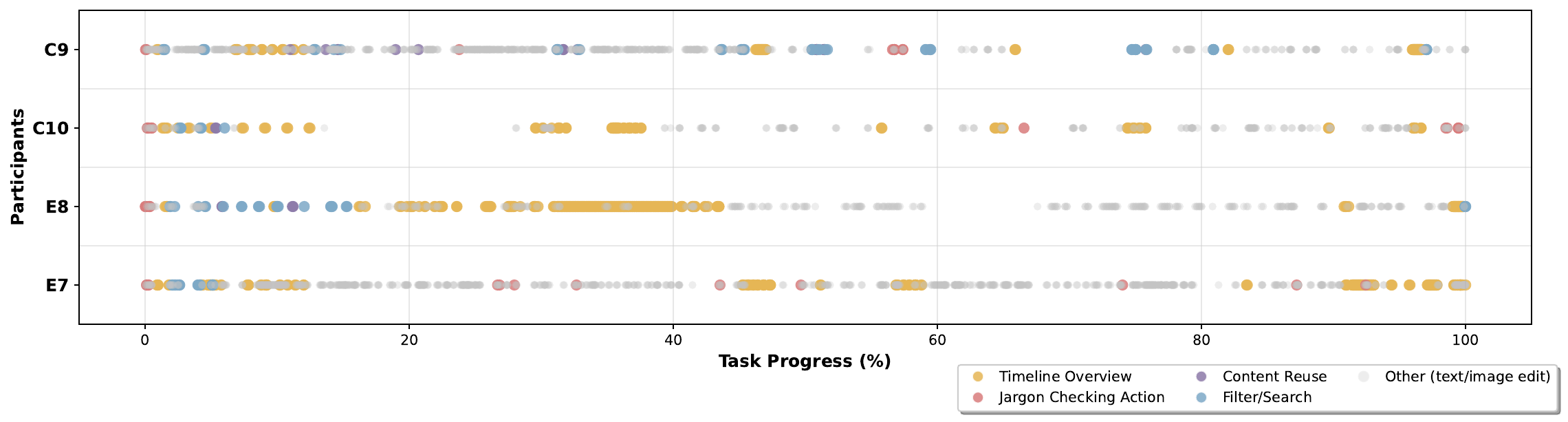}
    \vspace{-4mm}
    \caption{Four representative user-interaction sequences, selected from both the comparative and exploratory studies. These illustrate (1) different patterns of revisiting Timeline Overview, (2) different timing of jargon checking, and (3) different strategies of content reuse (exploratory vs. intentional). The Other category denotes residual actions such as text editing, image editing, or window switching.}
    \label{fig:interaction-log}
\end{figure*}

The Timeline Overview further supported this reasoning by combining \qt{importance of each part} (c1, c11) with its time allocation. 
Participants used it to \qt{stay mentally aware and structure slides within the time limit} (c1, c5, e3, e7) and \qt{break down the topic into subsections} (c8, c9).
Some participants admitted that, in Baseline, they unintentionally ignored the audience information, whereas explicitly setting the audience in \sysname{} kept this constraint visible and helped them \qt{estimate how many slides I should make} (e6) and how much emphasis each section should receive. 
Interaction logs confirmed the pattern, showing that multiple participants  (\autoref{fig:interaction-log}) set an initial timeline at the start, regularly returned to the overview during authoring, and performed final adjustments at the end to ensure their deliverable satisfied all constraints.

\subsection{Sections and Granularity Shifting Where Presenters Choose to Intervene (RQ2)}

Across both tools, participants described starting from \qt{a rough outline with high-level ideas} (c1, c4, c7, c10, e2, e3, e4, e6, e7), and perceived presentation authoring as connecting ideas, with each idea serving as a basic building block in their mental workflow (c10).
Baseline worked reasonably well when they already had a moderately defined storyline (c8), especially for ensuring smooth transitions.
However, it still centered their attention on individual slides, and participants emphasized that including all the necessary ideas in the presentation was just as important as polishing transitions (c4). 

With \sysname{}, some participants used Section to bridge their mental narratives and slides, deciding to intervene at the level of sections rather than just individual slides when structuring the story.  
They noted that Section helped them \qt{think from the section level, not just the slide level} (c12, e6), which \qt{matched their mental model} (c10, e8) and made the flow of high-level ideas visible (c1, c8, c12). 
Instead of keeping this structure in their heads, they could externalize it as an intermediate narrative granularity layer and then refine details within each section.

\subsubsection{The Timeline Overview Links Constraint Changes to Authoring Traces}
Defining sections with descriptions and later reorganizing them at both the section level and the slide level was described as a \qt{process of sorting out mental models of how the story should flow} (e3, e5, e7).
The Timeline Overview supported this process by making the section distributions visible during editing and helping participants track how ideas are constructed and flowed over time (c8, e5).
They appreciated having two separate views (e5), where one focused on structure (\autoref{fig:system}A) and the other focused on detailed content (\autoref{fig:system}C).
Because this setup made structurally complex presentations \qt{visually easier to comprehend}, than trying to keep the entire structure \qt{in [their] head} (c3, c5, c8, c12). 

In contrast, the absence of a clear layout and visual indicators in Baseline made navigation (c12) difficult, even when participants created workarounds such as \qt{manually defin[ing] the section slides} (e6, c12) or drafting outlines (c11, c12).
They therefore asked for more explicit \qt{multi-layered grouping} (e1, e5) that aligns with how they structure ideas across narrative levels. 

\subsubsection{Section-first Workflow Encouraged Narrative-centered Authoring}
Within \sysname{}'s section-first workflow, participants adopted different strategies when developing their narrative. 
Some followed a top-down (c1, e6, e5) approach, starting from \qt{high-level story chunks} then \qt{put details in} (c1) while later adjusting and reorganizing these chunks as their story evolved (c1).
Others used a bottom-up approach (c6, c12, e7), first listing all available ideas as sections (c12) and then organizing them into a coherent storyline, noting that \qt{Section actually comes very late in my presentations} (c6). 
In both cases, participants appreciated being able to externalize their initial ideas at a higher level and then revisit and reshape this structure as their narrative became clearer (e3, e4).
For the bottom-up workflows, the Timeline Overview acted less as a constraint but more as a surface that laid out existing ideas so that they could then reorganize based on their evolving storyline (c7, e2).


\subsection{Reuse, Versioning, and Cross-Session Narrative Traces (RQ3)}
\label{sec:theme3}

Participants found that \sysname{} effectively supported them in reusing slides.
While reuse was possible in Baseline, several participants ended up copying all existing slides into the workspace regardless of the content, then deleting those they did not need (c5, c6).
In contrast, they explicitly expressed their preferences towards the reuse in \sysname{} (c5, c6, c10). 
Interaction logs revealed three distinct strategies that informed their long-term workflows. (1) Exploratory reuse: participants browsed existing slides for ideas and inspiration (\autoref{fig:interaction-log}, c9, e7, e8). This was characterized by multiple filter actions without actual reuse, which is either concentrated at the start (e7, e8) or spread throughout the authoring cycle (c9). (2) Intentional reuse: participants adapted targeted content for the new presentation (\autoref{fig:interaction-log}, c9, c10, e8), typically involving multiple filtering actions followed by one or more reuse actions. (3) Versioning and synchronization: participants saved and aligned materials across multiple iterations to manage evolving content.

\subsubsection{Exploratory Reuse for Idea Generation and Inspiration}
\qt{Cold start} was a common problem (c5, c8, c12, e5) that reduced their authoring efficiency. 
Participants treat previous slides as a base of the new presentation, helping \qt{reducing the mental burden} (c5, e3, e4). 
This cold start could occur at the beginning of the entire presentation (c5, c8, c12, e4, e5) or when they lost inspiration for \qt{the middle sections} (c12). 
Depending on where they felt lost, they reused different units.
For the whole presentation, some participants were unsure about the content and simply added all existing slides before planning (c12, e3, e5), preferring the \qt{most recent presentations because they reflected more of my taste at the current moment} (e3). 
For the meat of a presentation, they tended to reuse specific sections that represented particular ideas, differentiating mainly in phrasing (c12). 
In other cases, they skimmed through relevant existing content only to \qt{look for idea collections}to inspire their own storyline and structure (c8, c9).

\subsubsection{Intentional Reuse for Targeted Content Adaptation}
In other situations, participants started with a rough mental model and reused content with a clear target. 
They examined whether the existing presentations were helpful in terms of effort required to adapt the content and favored those \qt{[they] didn't have to change much} (c3, e1). 
This intentional reuse happened at multiple granularity levels. 
At the narrative level, participants reused previous presentations or entire structures when the current storyline was \qt{exactly the same or similar to} what they had before (c6, e7, e8), then adapted diverging parts to fit their intended story flow and section allocations (e5, e7).
When the Timeline Overview indicated that storylines were very different, they instead started from scratch and reused only sections or slides (e2, e8). 
In these cases, participants began by adding one or more section(s) into the high-level storyline (c10, e3, e8), roughly configuring section parameters to match their mental model, then restructuring at a finer level by adding or adjusting slides (c10, e6, e7) or changing expressions to better fit the target audience (e6). 
At the slide level, reusable elements became more specific, such as introduction slides (c11), graphics (c9, c12), and textual components (e6).
Throughout authoring, participants regularly opened the Central Slide Repository to skim the previous slides but often used them as \qt{a guideline to double-check if what [they were] doing is correct} (c9) rather than copying them directly.

\subsubsection{Saving and Synchronizing Across Versions to Manage Evolving Content} 
Participants also used \sysname{} to save and synchronize slide content across the authoring process, again at multiple granularities. 
After finishing slides for one presentation, they intentionally saved content to the central slide repository for future reuse. 
The four provided options  (\ie, Ignore Changes, Save as Origin, Keep Both Versions, Replace Content) gave participants flexibility, but many found it difficult to decide which versions to replace. 
As a result, they often chose \qt{to keep both versions} as a \qt{safe fallback} (e5), or ignored changes when they recognized that edits were accidental, using the highlighted differences in the slide comparison window to make that decision.
Some participants also expressed their willingness to save sections (c10) and \qt{particular slides}, such as a core idea, introduction page, or conclusion page (e4), when they felt these were important and general enough for future reuse. 
These practices show how \sysname{} supports not only immediate reuse but also the gradual accumulation of cross-session narrative traces, as participants decide which versions and units to preserve as building blocks for subsequent presentations.
\section{Discussion}

\subsection{Reflections on the CMPA Framework}

Our work contributes CMPA, a conceptual framework that makes presenters' authoring rationale more tractable by linking three elements: contextual constraints, narrative granularity, and observable authoring traces. 
While we cannot directly access presenters' moment-to-moment reasoning, their edits and corresponding locations serve as external signals of internal decision-making \cite{zhang_representations_1994}. 
By associating these signals with the active constraint profile and narrative location, CMPA provides a reasoning layer that current tools and frameworks often miss \cite{roels_conceptual_2019, spicer_nextslideplease_2009,edge_slidespace_2016, takahashi_academic_2024}.
Therefore, tools can begin to inspect why a set of changes happened, and what kind of support would be useful when certain constraints are triggered, rather than relying on presenters to search for help after problems have already surfaced.

The CMPA framework is grounded in presenters' natural workflows, where presentation authoring is a multi-session and iterative process. 
Going beyond conceptual articulation, we instantiated CMPA through a research prototype and examined it with 20 participants, demonstrating how the framework can be operationalized and how it surfaces recurring user patterns in practice. 
Together, these insights suggested new opportunities for future presentation authoring tools to provide assistance that is more context-rooted, timely, and aligned with presenters' evolving constraints with narratives. 
At the same time, our studies only covered part of CMPA's design space.
They primarily covered how three representative contextual constraints evolve with corresponding authoring behaviors in a single-deck reuse scenario, leaving other kinds of constraints and long-term patterns for future work. 

CMPA also connects to broader human-AI co-creation frameworks by giving them high-level dimensions and domain-specific semantics in slide authoring. 
Within CMPA, phases in Moruzzi's User-Centered Co-Creativity (UCCC) framework, such as ``Feedback \& Adaptation,'' align with how contextual constraints initiate authoring, shape its execution, and guide evaluation of whether the resulting content fits the defined contextual constraints \cite{moruzzi_user-centered_2024}. 
UCCC's notions for user-defined rules and modalities correspond to CMPA's contextual constraints and narrative granularity. 
While both frameworks support user reflection on traces, CMPA extends this by adding a constraint-aware reasoning layer that links those traces to evolving contextual constraints.
In addition, CMPA provides a concrete representation of a shared, actionable context, and explicit intent signals emphasized in Co-Creative Framework for Interaction design (COFI) and Lin et al.'s Mixed-initiative co-creative (MI-CC) design space \cite{rezwana_designing_2023, lin_beyond_2023}.
In particular, contextual constraint values, narrative granularity levels, and revision intents specify what the presenter is trying to accomplish, at what scope, and when the initiation might shift between human and assistant.
Taken together, these connections position CMPA as a constraint-aware narrative layer that co-creative slide authoring assistants can track and manipulate.

Beyond our prototype, CMPA can also be instantiated in other ways during presentation authoring.
For example, it could be used as an analytics layer that segments users' interaction logs into different constraint-driven episodes to support reflection and discussion, or drive a co-creative assistant that proposes real-time edit suggestions at appropriate narrative granularities when specific constraints shift. 
In this sense, CMPA offers a foundation that other researchers can build upon by extending the constraint set, applying it to different media, or integrating it with human-AI co-creation workflows. 



\subsection{Design Implications}

\subsubsection{Future Tools Should Support Constraint-Aware Versioning Beyond Content Reuse}
Future presentation tools should support a constraint-aware layer for versioning and reuse, extending reuse from chronological, pure content-level operation to narrative-level reasoning. 
Existing approaches focus on identifying semantically similar slide fragments and enabling reuse across multiple variants \cite{roels_conceptual_2019, spicer_nextslideplease_2009}. 
However, these variants are often disconnected from the contextual constraints that motivated them, making it difficult to understand why an alternative exists and when it should be reused.
\rev{
For example, a presenter may want to further simplify a concept to better match the audience's background knowledge. In this case, the relevant slides are variations created to accommodate a specific audience group. When multiple versions of similar content exist for different purposes, such as time limits, audience expertise, or communicative intent, similarity-based retrieval still requires presenters to manually infer which version best fits their current goal. 
}
Moreover, CMPA suggests that shifts in contextual constraints frequently trigger coherent authoring behaviors that realign the narrative with the new context.
This implies that presentation versioning should be organized around both what content change and what associated contextual conditions drove the change.
In practice, future systems should enable authors to retrieve prior presentation versions by the contextual constraints in which they were created (e.g., selecting versions aligned with a given contextual constraint value) and capture lightweight information about the author's intent or behaviors during revision, reducing reliance on memory and file metadata during reuse.


\subsubsection{Future Tools Should Carefully Balance Between AI Automation and Author Controls}
Our findings also show that audience-oriented support is more effective when embedded into the authoring process rather than offered as post-hoc repair. 
In \sysname{}, jargon checking flagged difficult terms and acted as a reflective trigger that encouraged presenters to refine explanations, anticipate questions, and adjust expressions for different audiences. 
While some participants appreciated jargon checking as an on-demand interaction (c7, c10, e8), others wanted it to operate more proactively and continuously during authoring (e5, e7). 
This points to a broader design implication for mixed-initiative authoring: future systems should balance proactive suggestions with the author's agency and sense of control  \cite{hoque_hallmark_2024,hutson_human-ai_2025,reza_co-writing_2025}. 
Prior work highlights the importance of making control options explicit and supporting based on context \cite{chen_need_2024, wang_investigating_2024}
In the context of constraint-driven presentation authoring, such a balance is particularly important because presenters are making nuanced tradeoffs across contextual constraints. 
Future tools should therefore support timely automation and reflection while preserving authors' ability to make decisions, adjust narrative granularity levels, and maintain ownership over options.

\subsection{Limitations and Future Work}
Our work has several limitations. 
First, our study was conducted in a lab setting with limited time. This setting enabled us to observe how contextual constraints, narrative revision intent, and authoring behaviors manifest in a controlled way, but it underrepresented the longer and messier aspects that our CMPA framework is designed to describe, especially the accumulation of multi-session narrative traces.
Synchronization behaviors were only examined in the second session of the exploratory study, and we limited available slide authoring functionalities (\ie shapes, animations, and audio), to focus on core interactions. 
\rev{
Similarly, multi-session reuse was evaluated through a controlled two-session simulation, lacking evidence from longer-term real-world use.
}
Longer-term use with richer authoring functionality and more diverse presenters (\ie novices, industrial workers, other disciplines) may reveal different constraint profiles, granularities of change, and nuanced reuse strategies than we observed. 
\rev{
Future work should therefore deploy \sysname{} in more realistic settings to better understand how cross-session narratives accumulate and evolve.
}

Second, \sysname{} and our studies only partially instantiated a subset of the design space implied by CMPA. 
We have not yet modeled other important aspects of presentation context, such as venue, technical setup, or institutional policies, nor have we studied collaboration or industrial settings. 
Likewise, we did not store contextual constraint profiles across sessions, and our jargon checking only considers expertise level, which does not fully capture other important audience characteristics. also relies on a general-purpose language model and currently supports only English slides. 
Although the underlying pattern of the CMPA may extend these richer contexts and constraints, we have not empirically validated these extensions. 
These choices are appropriate for a research probe, but future work should investigate additional contextual constraints, deploy \sysname{} over longer periods, and use them to refine or extend the CMPA.

\rev{
Third, our comparison to baseline was limited.
The baseline tool included only basic editing and copying functions, whereas mainstream presentation tools often provide AI-assisted features and access to external libraries.
We intentionally chose a minimal baseline to better control for familiarity, but this choice likely under-represented the capabilities of real-world slide authoring tools and therefore limited the external validity of the comparison. 
Future work could compare \sysname{} against stronger and more realistic baselines to better assess both its practical benefits and the applicability of the CMPA framework in commercial presentation authoring contexts.

Fourth, our work focuses on slide authoring behaviors within and across sessions, and does not explicitly model higher-level structures beyond presentation-level authoring. 
We have not yet captured earlier and broader stages of presentation authoring, such as ideation or delivery, nor have we modeled higher-level factors such as user-models, thematic influences, or usage scenarios that may shape how presenters differentiate scenarios or frame topics before producing presentations.
Although the underlying logic of CMPA may extend to these higher-level structures and influences, we have not yet empirically validated these extensions. 
These choices are appropriate for a research probe, but future work should extend CMPA to account for richer stages of authoring cycle and broader factors that influence slide design decisions. 
}
\section{Conclusion}
\rev{
In this paper, we introduced CMPA, a conceptual framework that systematically organizes contextual constraints from isolated components into continuous driving forces of how presentation narratives are revised and reused over multiple authoring sessions.
}
We instantiated CMPA through \sysname{}, a research prototype that integrates contextual constraints into the iterative presentation authoring process through Section, a unit for grouping slides based on ideas, together with Timeline Overview, Slide Editor, and a centralized library, enabling flexible reuse at multiple granularities. 
\rev{
Through a comparative within-subject study and an two-session exploratory study, we showed how CMPA can help presenters better align evolving presentation content with shifting contextual constraints across multiple authoring sessions.
}
The design implications derived from this work shed light on future authoring environments that more explicitly support iterative, constraint-driven content creation and reuse.

\begin{acks} 
This work is supported in part by the Natural Sciences and Engineering Research Council of Canada (NSERC) Discovery Grant \#RGPIN-2020-03966 and the Ontario Early Researcher Award (ERA) \#ER24-18-222, and the Canada Foundation for Innovation (CFI) John R. Evans Leaders Fund (JELF) \#42371.
\end{acks}

\bibliographystyle{ACM-Reference-Format}
\bibliography{references.bib}

\newpage
\appendix
\appendix
\section{Formative Study Details} \label{apx:formative}

\subsection{Demographic information for participants}

\begin{table*}[tb]
\centering
\caption{Demographic information for formative participants}
\begin{tabular}{lllll}
\toprule
\textbf{Participant ID} & \textbf{Age} & \textbf{Gender} & \textbf{Occupation} & \textbf{Field of Study}\\
\midrule
P1  & 34 & Man & Student, PhD & Information Visualization\\
P2  & 34 & Man & Postdoctoral researcher & HCI/Design/AIoT\\
P3  & 26 & Woman & Student, PhD & Human Computer Interaction\\
P4  & 25 & Woman & Student, PhD & Human Computer Interaction\\
P5  & 29 & Woman & Student, PhD & Human Computer Interaction\\
P6  & 28 & Man   & Student, PhD & Human Computer Interaction\\
P7  & 22 & Woman & Student, PhD & Computer Science\\
P8  & 25 & Woman   & Student, PhD & Computer Science\\
P9  & 24 & Man & Student, PhD & VR/AR/Haptics\\
P10 & 25 & Man & Student, PhD & Programming Interface\\
\bottomrule
\end{tabular}
\end{table*}

\section{Jargon Checking prompts}
\label{apx:jargonPrompts}

\subsection{Prompts for Audience Context Expansion}

\begin{lstlisting}[basicstyle=\ttfamily\small,breaklines=true,
caption={Prompt for audience context expansion used in ReSlide},
label={lst:audiencePrompt}]
You are an expert at understanding audience descriptions for presentations.
Analyze this audience description and provide detailed context that will help with jargon detection.

ORIGINAL AUDIENCE DESCRIPTION: "${originalDescription}"
USER-PROVIDED EXPERTISE LEVEL: ${userExpertiseLevel}/5

Your task: Expand this into a detailed profile that clearly explains what this audience would and would not know.

Respond in JSON format:
{
  "expandedDescription": "Detailed 2-3 sentence description of their background and knowledge level",
  "inferredExpertiseLevel": number (1-5, can adjust user's estimate if clearly wrong),
  "knownConcepts": ["concept1", "concept2", "concept3"],
  "likelyJargon": ["term1", "term2", "term3"],
  "domainBackground": "Their field/industry background"
}

EXAMPLES:

Input: "NLP professor"
Output: {
  "expandedDescription": "Computer Science professor specializing in Natural Language Processing research. Has PhD-level expertise in machine learning, deep learning, linguistics, and computational methods. Familiar with all standard ML algorithms, programming concepts, and academic research terminology.",
  "inferredExpertiseLevel": 5,
  "knownConcepts": ["machine learning", "neural networks", "transformers", "BERT", "random forest", "SVM", "tokenization", "embeddings", "algorithms", "APIs", "frameworks", "deep learning", "statistical models"],
  "likelyJargon": ["novel architectures from 2024", "proprietary model names", "company-specific tools", "bleeding-edge research terms"],
  "domainBackground": "Computer Science academia with focus on NLP/AI research"
}

Input: "undergrad freshman no programming experience"
Output: {
  "expandedDescription": "First-year undergraduate student with no prior programming or computer science background. Familiar with basic technology use (smartphones, apps, social media) but unfamiliar with technical concepts, programming terminology, or how software systems work.",
  "inferredExpertiseLevel": 1,
  "knownConcepts": ["apps", "websites", "social media", "AI tools like ChatGPT", "smartphones", "basic internet concepts"],
  "likelyJargon": ["programming terms", "algorithms", "databases", "machine learning", "neural networks", "APIs", "coding concepts"],
  "domainBackground": "General education, non-technical"
}

Be specific about what they would know vs. what would be jargon.
Focus on their domain expertise.
\end{lstlisting}

\subsection{Prompt for Jargon Detection}

\begin{lstlisting}[basicstyle=\ttfamily\small, breaklines=true, caption={Prompt for jargon detection used in ReSlide}, label={lst:jargonPrompt}]
Analyze this slide content for jargon terms that would confuse the specified audience.

AUDIENCE PROFILE:
- Original Description: "${expandedContext.originalDescription}"
- Detailed Profile: ${expandedContext.expandedDescription}
- Expertise Level: ${expandedContext.inferredExpertiseLevel}/5
- Domain Background: ${expandedContext.domainBackground}
${presentationContext ? `- Presentation Context: ${presentationContext}` : ''}

WHAT THIS AUDIENCE KNOWS:
${expandedContext.knownConcepts.length > 0
            ? expandedContext.knownConcepts.map(concept => `- ${concept}`).join('\n')
            : '- No specific known concepts provided'}

LIKELY JARGON FOR THIS AUDIENCE:
${expandedContext.likelyJargon.length > 0
            ? expandedContext.likelyJargon.map(jargon => `- ${jargon}`).join('\n')
            : '- No specific jargon areas identified'}

SLIDE CONTENT TO ANALYZE:
Title: ${slideTitle || 'Untitled'}
Content: ${slideText}

CRITICAL INSTRUCTIONS:
1. Use the detailed audience profile to determine what would be jargon
2. If a term is in the "WHAT THIS AUDIENCE KNOWS" list, it's NOT jargon
3. If a term is similar to items in "LIKELY JARGON" list, it probably IS jargon
4. Consider the expertise level (${expandedContext.inferredExpertiseLevel}/5) carefully
5. Only flag terms that would genuinely prevent understanding

EXPERTISE LEVEL GUIDELINES:
- Level 1-2: Most technical terms are jargon, but common tech words (AI, app, website) are okay
- Level 3: Specialized and domain-specific terms are jargon
- Level 4: Only cutting-edge or highly specialized terms are jargon
- Level 5: Only the most novel, bleeding-edge, or extremely specialized terms are jargon

IMPORTANT: 
- For professors/experts (Level 4-5): Very few terms should be jargon
- For beginners (Level 1-2): Many technical terms will be jargon
- Focus on the audience's specific domain background: ${expandedContext.domainBackground}

Respond with JSON only (no markdown):
{
  "jargonTerms": [
    {
      "term": "exact term from text",
      "definition": "Clear explanation appropriate for this audience",
      "alternatives": ["simpler alternative 1", "accessible phrase 2"],
      "startIndex": number,
      "endIndex": number
    }
  ]
}
\end{lstlisting}

\section{Evaluation Study Details} \label{apx:evaluation}
\subsection{Screening Questions}
\subsubsection{Definition for \textbf{presentation} for the following screening questions:}

\begin{itemize}
    \item Included \textbf{planned content}
    \item Had \textbf{a clear and purposeful structure} 
    \item Used \textbf{visual aids} (like slides)
    \item Were \textbf{tailored to a specific audience},  \textbf{setting}, \textbf{time limit}, or \textbf{goal}.
\end{itemize}

\subsubsection{Presentation skill assessment}
\begin{enumerate}
    \item How often did you give presentations in the past 3 months?
    \begin{itemize}
        \item 0
        \item 1-2 times
        \item 3-5 times
        \item 6-10 times
        \item More than 10 times
    \end{itemize}

    \item How confident are you in your overall presentation experiences? 
    \begin{itemize}
        \item Not at all confident
        \item Slightly confident
        \item Moderately confident
        \item Very confident
        \item Extremely confident
    \end{itemize}
    \item Have you received any formal training in public speaking or presentation skills? 
     \begin{itemize}
        \item No
        \item Yes, informal trainings (e.g. workshops, guides, or feedback from others)
        \item Yes, formal instructions (e.g. a course, debate/public speaking club)
        \item Other:
    \end{itemize}
    \item About how many years have you been giving presentations (e.g. in classes, workshops, or conferences) since post-secondary education?
    \item Which best describes the typical audience for your past presentations ? (Choose the most common)
    \begin{itemize}
        \item Mostly peers or classmates
        \item Mixed audiences (e.g. peers, instructors, or professionals)
        \item Primarily professional audiences (e.g. clients, conferences, external stakeholders)
        \item Other:
    \end{itemize}
    \item I have presented to a large audience (e.g., more than 50 people).
    \begin{itemize}
        \item Never true
        \item Rarely true
        \item Sometimes but infrequently true
        \item Neutral
        \item Sometimes true
        \item Usually true
        \item Always true
    \end{itemize}
    \item How often did you use slide creation software to prepare presentations?
    \begin{itemize}
        \item Never
        \item Rarely
        \item  Sometimes
        \item Often
        \item Always
    \end{itemize}
    \item Which slide creation tools were you familiar with? (Select all that apply)
    \begin{itemize}
        \item Microsoft Power Point
        \item Google Slides
        \item Apple Keynote
        \item Prezi
        \item Canva
        \item Other
    \end{itemize}

\end{enumerate}

\subsection{Demographic information for within-subject participants}

\begin{table}[tbp]
\centering
\caption{Demographic information for within-subject participants}
\begin{tabular}{lllp{3cm}}
\toprule
\textbf{Participant ID} & \textbf{Age} & \textbf{Gender} & \textbf{Occupation or Field of Study} \\
\midrule
C1  & 25 & Woman & Software Developer \\
C2  & 24 & Woman & NLP \\
C3  & 33 & Woman & Data Science \\
C4  & 32 & Woman & HCI--AI \\
C5  & 22 & Woman & Student, PhD \\
C6  & 23 & Man   & Computer Science \\
C7  & 24 & Woman & Computer Science \\
C8  & 24 & Man   & Computer Science \\
C9  & 25 & Woman & Student, Master's \\
C10 & 24 & Woman & Student, PhD \\
C11 & 25 & Man   & Student, PhD \\
C12 & 26 & Woman & Data Science \\
\bottomrule
\end{tabular}
\end{table}

\subsection{Demographic information for exploratory participants}

\begin{table}[tbp]
\centering
\caption{Demographic information for exploratory participants}
\begin{tabular}{lllp{3cm}}
\toprule
\textbf{Participant ID} & \textbf{Age} & \textbf{Gender} & \textbf{Occupation or Field of Study} \\
\midrule
E1  & 25 & Man & Computer Science \\
E2  & 39 & Man & Data Science \\
E3  & 28 & Man & Computer Science \\
E4  & 34 & Woman & Studied Finance \\
E5  & 34 & Man & HCI \\
E6  & 25 & Woman   & Computer Science \\
E7  & 25 & Woman & Computer Science \\
E8  & 29 & Woman   & Computer Science \\

\bottomrule
\end{tabular}
\end{table}

\subsection{Custom Likert Scale Questions}
\subsubsection{Contextual Constraints related}
\begin{enumerate}
    \item I felt the slides I created captured constraints for time, audience, and presentation intent.
    \item The system helped me stay mentally aware and structured my slides within the time limit.
    \item The system helped me map my communicative intent to the presentation content. 
    \item The system increased my awareness of matching language to the target audience’s needs.
    \item The system helped me express ideas in a way that’s clear and relevant to my audience.
\end{enumerate}

\subsubsection{Narrative Coherence}
\begin{enumerate}
    \item The system helped me structure and organize my thoughts while making the presentation.
    \item I found it easy to plan my overall presentation flow.
    \item I found it easy to switch between/adjust both the overall structure and detailed content organization during authoring.
\end{enumerate}

\subsubsection{Presentation Content Reuse}
\begin{enumerate}
    \item I found it convenient and efficient to reuse slides using this system. 
    \item It was easy to understand how to reuse slides with the system.
    \item The system gave me control over how my existing slides were reused and updated. 
\end{enumerate}

\end{document}